\documentclass[a4paper,twocolumn,11pt,accepted=2025-03-27]{quantumarticle}
\pdfoutput=1
\usepackage[utf8]{inputenc}
\usepackage[english]{babel}
\usepackage[T1]{fontenc}

\usepackage{color,amsthm,amsmath,amsxtra,amsfonts,dsfont,graphicx,bm,amssymb}
\usepackage[colorlinks=true,linkcolor=blue, citecolor=blue, urlcolor=blue, bookmarks]{hyperref}
\usepackage{centernot}
\usepackage[dvipsnames]{xcolor}
\usepackage{graphicx}
\usepackage{tikz}
\usepackage{braket}
\usepackage{multirow}
\usepackage{cite}

\newcommand{\be}{\begin{equation}}
	\newcommand{\ee}{\end{equation}}
\newcommand{\bea}{\begin{eqnarray}}
	\newcommand{\eea}{\end{eqnarray}}
\newcommand{\bse}{\begin{subequations}}
	\newcommand{\ese}{\end{subequations}}

\theoremstyle{plain}

\makeatletter
\DeclareFontFamily{OMX}{MnSymbolE}{}
\DeclareSymbolFont{MnLargeSymbols}{OMX}{MnSymbolE}{m}{n}
\SetSymbolFont{MnLargeSymbols}{bold}{OMX}{MnSymbolE}{b}{n}
\DeclareFontShape{OMX}{MnSymbolE}{m}{n}{
    <-6>  MnSymbolE5
   <6-7>  MnSymbolE6
   <7-8>  MnSymbolE7
   <8-9>  MnSymbolE8
   <9-10> MnSymbolE9
  <10-12> MnSymbolE10
  <12->   MnSymbolE12
}{}
\DeclareFontShape{OMX}{MnSymbolE}{b}{n}{
    <-6>  MnSymbolE-Bold5
   <6-7>  MnSymbolE-Bold6
   <7-8>  MnSymbolE-Bold7
   <8-9>  MnSymbolE-Bold8
   <9-10> MnSymbolE-Bold9
  <10-12> MnSymbolE-Bold10
  <12->   MnSymbolE-Bold12
}{}

\let\llangle\@undefined
\let\rrangle\@undefined
\DeclareMathDelimiter{\llangle}{\mathopen}%
                     {MnLargeSymbols}{'164}{MnLargeSymbols}{'164}
\DeclareMathDelimiter{\rrangle}{\mathclose}%
                     {MnLargeSymbols}{'171}{MnLargeSymbols}{'171}
\makeatother

\begin{document}
	
	\title{Approximate inverse measurement channel for shallow shadows}
	
	\author{Riccardo Cioli}
	\affiliation{Dipartimento di Fisica e Astronomia, Università di Bologna and INFN, Sezione di Bologna, via Irnerio 46, I-40126 Bologna, Italy}
	
	\author{Elisa Ercolessi}
	\affiliation{Dipartimento di Fisica e Astronomia, Università di Bologna and INFN, Sezione di Bologna, via Irnerio 46, I-40126 Bologna, Italy}
	
	\author{Matteo Ippoliti}
	\affiliation{Department of Physics, The University of Texas at Austin, Austin, TX 78712, USA}
	
	\author{Xhek Turkeshi}
	\affiliation{Institut für Theoretische Physik, Universität zu Köln, Zülpicher Strasse 77a, 50937 Köln, Germany}
	
	\author{Lorenzo Piroli}
	\affiliation{Dipartimento di Fisica e Astronomia, Università di Bologna and INFN, Sezione di Bologna, via Irnerio 46, I-40126 Bologna, Italy}
	
\begin{abstract}

Classical shadows are a versatile tool to probe many-body quantum systems, consisting of a combination of randomised measurements and classical post-processing computations. In a recently introduced version of the protocol, the randomization step is performed via unitary circuits of variable depth $t$, defining the so-called shallow shadows. For sufficiently large $t$, this approach allows one to get around the use of non-local unitaries to probe global properties such as the fidelity with respect to a target state or the purity. Still, shallow shadows involve the inversion of a many-body map, the measurement channel, which requires non-trivial computations in the post-processing step, thus limiting its applicability when the number of qubits $N$ is large. In this work, we put forward a simple approximate post-processing scheme where the infinite-depth inverse channel is applied to the finite-depth classical shadows and study its performance for fidelity and purity estimation. The scheme allows for different circuit connectivity, as we illustrate for geometrically local circuits in one and two spatial dimensions and geometrically non-local circuits made of two-qubit gates.  For the fidelity, we find that the resulting estimator coincides with a known linear cross-entropy, achieving an arbitrary small approximation error $\delta$ at depth $t=O(\log (N/\delta))$ (independent of the circuit connectivity). For the purity, we show that the estimator becomes accurate at a depth $O(N)$. In addition, at those depths, the variances of both the fidelity and purity estimators display the same scaling with $N$ as in the case of global random unitaries. We establish these bounds by analytic arguments and extensive numerical computations in several cases of interest. Our work extends the applicability of shallow shadows to large system sizes and general circuit connectivity.
\end{abstract}

	\maketitle
	
	

\section{Introduction}  

 The successful implementation of many ideas in quantum information science depends on our ability to probe certain properties of many-body quantum states efficiently. Yet, given a large quantum system of $N$ qubits, performing complete tomography of its state is often unfeasible, requiring a number of measurements scaling exponentially in $N$~\cite{flammia2012quantum,haah2016sample,o2016efficient}. As larger devices and platforms become available within existing quantum technology~\cite{blatt2012quantum,gross2017quantum,preskill2018quantum,schafer2020tools,kjaergaard2020superconducting,morgado2021quantum,monroe2021programmable,alexeev2021quantum,pelucchi2022potential,burkard2023semiconductor}, it is thus very important to find efficient protocols to probe selected properties of many-body states, without making use of full-state tomography.

Much recent progress in this direction has followed the development of the so-called randomised-measurement (RM) toolbox~\cite{elben2019statistsical, huang2020predicting, elben2023randomized,cieslinski2023analysing}. A prominent element in this toolbox is the classical-shadow approach~\cite{huang2020predicting}. It consists of applying random unitary operators, drawn from some suitable ensemble, to each copy of an unknown quantum state $\rho$, which is then measured on the computational basis. Notably, the measurements need not be tailored to a specific property of the system. Rather, the measurement outcomes can be processed to build estimators for different quantities of interest~\cite{vanEnk2012measuring,elben2018renyi,knips2020multipartite,ketterer2019characterizing}.

The ensemble of random unitaries determines the number of measurements needed to estimate a given quantity accurately~\cite{elben2023randomized}. The two original versions of the protocol~\cite{huang2020predicting} involve either random single-qubit rotations, defining the so-called random Pauli measurements, or global unitaries drawn randomly from the set of Clifford operators~\cite{gottesman1997stabilizer}. While the former is suited to probe local properties such as few-qubit correlations~\cite{huang2020predicting,ippoliti2024classical}, the use of global Clifford unitaries outperforms the local Pauli measurements to estimate global quantities. Two important examples are the fidelity with respect to a given target state, an essential building block for quantum-state certification~\cite{kliesch2021theory,carrasco2021theoretical,daSilva2011practical,flammia2011direct,aolita2015reliable,gluza2018fidelity,takeuchi2018verification}, and the purity~\cite{nielsen2002quantum}. Unfortunately, realizing global unitaries is nontrivial in practice and represents a bottleneck for current noisy intermediate-scale quantum (NISQ) devices~\cite{preskill2018quantum}. 

From the experimental point of view, a less demanding protocol can be obtained by replacing the global operators with quantum circuits made of random two-qubit gates, yielding the so-called shallow shadows~\cite{bertoni2022shallow,akhtar2023scalable,arienzo2023closed,ippoliti2023operator}. For sufficiently large depth $t$, quantum circuits allow one to get around the use of global unitaries to estimate global quantities, making them appealing for current experimental implementation~\cite{hu2024demonstration,farias2024robust}. For instance, it was shown that this scheme allows us to accurately estimate the fidelity with $O(1)$ measurements when the circuit depth scales as $O(\log N)$~\cite{bertoni2022shallow}. Yet, the protocol requires non-trivial computations at the post-processing step, as one needs to invert a many-body map, the \emph{measurement channel}. While efficient tensor-network-based algorithms for this task were provided for shallow one-dimensional (1D) circuits~\cite{bertoni2022shallow,akhtar2023scalable}, their implementation is technically nontrivial, and no efficient algorithms are known beyond the 1D architecture, limiting the method's applicability. 
Other strategies to overcome the inversion of the measurement channel have been developed, see for example Refs.~\cite{malmi2024enhanced, mangini2024lowvariance}.

In this work, we put forward an approximate inversion of the measurement channel, consisting of applying the infinite-depth inverse channel to the finite-depth classical shadows. We focus on the problem of estimating the global properties of the system, for which global Cliffords provide an advantage over random Pauli measurements, studying, in particular, the fidelity and the purity. The scheme involves simple post-processing computations and allows for different circuit connectivity, as we illustrate for geometrically local circuits in one and two spatial dimensions and geometrically non-local circuits made of two-qubit gates. 

For the fidelity, we find that the resulting estimator coincides with a known linear cross entropy~\cite{arute2019quantum,neill2018blueprint}, achieving an arbitrary small approximation error $\delta$ at depth $t=O(\log (N/\delta))$ -- for all the circuit connectivities we considered. For the purity, we show that the estimator becomes accurate at a depth $O(N)$. In addition, the estimator variance for both the fidelity and purity at those depths displays the same scaling with $N$ as that obtained via global random Cliffords. We establish these bounds by analytic arguments and extensive numerical computations in several cases of interest. Our results extend the applicability of shallow shadows to large system sizes and general circuit connectivity, with potential practical applications in current quantum computing platforms.

The rest of this work is organised as follows. We begin in Sec.~\ref{sec:standard_quantum_shadows} by reviewing classical shadows based on shallow circuits and presenting the approximate inversion formula. In Sec.~\ref{sec:approximate_estimator_fidelity}, we study the resulting approximate fidelity estimator and the corresponding quantum-certification protocol, while in Sec.~\ref{sec:approximate_purity_estimator}, we report our analysis of the approximate purity estimator. We summarize our conclusions and outline directions for future work in Sec.~\ref{sec:outlook}. The most technical aspects of our work are consigned to several appendices. 

\section{Approximate inversion formula} 
\label{sec:standard_quantum_shadows}

\subsection{Classical shadows and the measurement channel}

This section recalls some aspects of classical shadows used in our work. We consider a system of $N$ qubits and denote by $\ket{0}_j$, $\ket{1}_j$ the elements in the local computational basis associated with spin $j$.  Given $M$ sequential experimental runs preparing a copy of $\rho$, we apply to each of them a random unitary operator sampled from a suitable ensemble $\mathcal{U}$, followed by a projective measurement in the computational basis $|b\rangle=\otimes_{j=1}^N \ket{b_j}_j$, with $b_j=0,1$. We will consider the ensemble of quantum circuits made of two-qubit Clifford gates. Namely, we will choose $U_t=V_t V_{t-1} \cdots V_1$, where $t$ is the circuit depth, while each ``layer'' $V_j$ contains quantum gates acting on disjoint pairs of qubits. Each gate is drawn from the uniform distribution on the set of two-qubit Clifford unitaries~\cite{gottesman1997stabilizer,gottesman1998theory}. 
 
    We consider two circuit architectures: geometrically local brickwork structures both in 1D and 2D and geometrically non-local structures. The former case consists of gates pairing qubits which are nearest neighbors on a regular $D$-dimensional lattice. For $D=1$, the circuit alternates layers of gates acting on pairs of qubits at positions $(j,j+1)$, with $j$ even and odd. For $D=2$, we choose the same connectivity of Ref.~\cite{sierant2023entanglement}, where each layer contains gates that pair a qubit with one of its four neighbors, cycling through the four neighbors over four layers. The geometrically non-local quantum circuits are instead constructed as follows: at each time step, we randomly partition the set of qubits into disjoint pairs and apply a random gate to each pair\footnote{These circuits are sometimes called Brownian random circuits~\cite{gharibyan2018onset,piroli2020random}}.
	
	For each run of the experiment, one constructs a classical ``snapshot'' of the quantum state, $U_t^\dagger \ket{b}$, based on the choice of unitary $U_t$ and the measurement outcome $b\in\{0,1\}^N$. Averaging over such snapshots yields the measurement channel 
	\begin{equation}\label{eq:measurement_channel}
	\mathcal{M}_t(\rho)=\underset{U_t \sim \mathcal{U}}{\mathbb{E}}\left[\sum_{b \in\{0,1\}^N}\left\langle b\left|U_t \rho U_t^{\dagger}\right| b\right\rangle U_t^{\dagger}|b\rangle\langle b| U_t\right]\,.
	\end{equation}
    Assuming knowledge of $\mathcal{M}_t$ and (crucially) of its inverse $\mathcal{M}_t^{-1}$, one can then define the classical shadows
\begin{equation}
	\label{eq:global_classical_shadow}
	\rho^{(r)}_t=\mathcal{M}_t^{-1}\left[(U_t^{(r)})^{\dagger}|b^{(r)}\rangle\langle b^{(r)}| U^{(r)}_t\right]\,,
 \end{equation}
 while $r\in\{1,\dots M\}$ labels the experimental runs. 
 
Classical shadows are stored and processed classically to obtain an estimate of different quantities of interest. As mentioned, we will focus on two global quantities. First, we  consider the fidelity with respect to a pure target state $\ket{\psi}$, for which we construct the estimator
\begin{equation}\label{eq:fidelity_estimator}
\mathcal{F}^{(e)}_t=\frac{1}{M}\sum_{r=1}^{M}f_t^{(r)},
\end{equation}
where 
\begin{equation}
 f_t^{(r)}=\langle{\psi} | \rho^{(r)}_t|\psi\rangle\,.
\end{equation}
Second, we  study the purity of the system state $\rho$, corresponding to the estimator
\begin{equation}\label{eq:estimator_purity}
\mathcal{P}_t^{(e)}=\frac{1}{M(M-1)} \sum_{r \neq r^{\prime}} {\rm Tr}\left(\rho_{t}^{(r)} \rho_{t}^{\left(r^{\prime}\right)}\right)\,.
\end{equation}
Using the definition of $\mathcal{M}_t$ in Eq.~\eqref{eq:measurement_channel}, it is straightforward to see that $\mathcal{F}_t^{(e)}$ and $\mathcal{P}_t^{(e)}$ are faithful, namely 
\begin{align}
{\mathbb{E}}[\mathcal{F}_t^{(e)}]&=\mathcal{F}_\psi(\rho)=\braket{\psi| \rho |\psi}\,,\\
{\mathbb{E}}[\mathcal{P}_t^{(e)}]&=\mathcal{P}(\rho)={\rm Tr} \left[\rho^2 \right]\,,
\end{align}
where the average is over the random unitaries and the projective measurement outcomes. 
	
The value of $M$ guaranteeing a small deviation of $\mathcal{F}^{(e)}_t$ from $\mathcal{F}_\psi(\rho)$ is determined by the variance of the random variable $f^{(r)}$. In turn, the latter is bounded by the square of the \emph{shadow norm} $||O||_{\rm sh}$~\cite{huang2020predicting} of the traceless operator 
 \begin{equation}\label{eq:o_operator}
 O=\ket{\psi}\bra{\psi}-\frac{\openone}{2^{N}}\,,
 \end{equation}
 which reads
	\begin{align}
		||O||^2_{\rm sh}=\underset{\rho\ {\rm state}}{\max}\left\{\underset{U_t}{\mathbb{E}}\sum_{_{b \in\{0,1\}^N}}\left[\left\langle b\left|U_t \rho U_t^{\dagger}\right| b\right\rangle\right.\right.\nonumber\\
		\left.\left.\times \left\langle b\left|U_t \mathcal{M}_t^{-1}(O) U_t^{\dagger}\right| b\right\rangle^2\right]\right\}\,.\label{eq:shadow_norm}
	\end{align}
	Fixing $O$ as in Eq.~\eqref{eq:o_operator}, the shadow norm depends on the unitary ensemble $\mathcal{U}$: $||O||_{\rm sh}^2$ grows exponentially in $N$ in the case of local Pauli measurements, while $||O||_{\rm sh}^2=O(1)$ for global Clifford unitaries~\cite{huang2020predicting}. Finite-depth circuits interpolate between these two extremes, recovering the global Clifford bound for $t\to\infty$. In fact, both numerical and analytical evidence show that the shadow norm associated with the fidelity estimator becomes $O(1)$ already at ``shallow depth'', namely $t\sim \log N $~\cite{bertoni2022shallow,akhtar2023scalable,arienzo2023closed}.

Similarly, the number of measurements needed to achieve an accurate description of the purity depends on the variance of the estimator $\mathcal{P}^{(e)}_t$. In this case, explicit bounds are known for the two limiting cases of Pauli measurements (corresponding to $t=0$) and global Cliffords ($t=\infty)$. In the former case, we have~\cite{elben2018renyi,brydges2019probing,elben2019statistsical}
\begin{align}\label{eq:variance_purity_pauli}
	{\rm Var}\left[\mathcal{P}_0^{(e)}\right]&\leq 4\left(\frac{2^{N}\mathcal{P}(\rho)}{M}\right)
	+2\left(\frac{2^{2N}}{M-1}\right)^2\,,
\end{align}
while in the latter case~\cite{huang2020predicting}
\begin{align}\label{eq:variance_purity_global}
	{\rm Var}\left[\mathcal{P}^{(e)}_\infty\right]&\leq \left(\frac{12\mathcal{P}(\rho)}{M}\right)
	+\frac{2\left(9\times 2^{2N}\right)}{(M-1)^2}\,,
\end{align}
where we have neglected terms that are exponentially small in $N$. We see that the bound on the variance of $\mathcal{P}_\infty^{(e)}$ grows exponentially in $N$ even for global random Cliffords. Still, the leading term for a fixed $M$ grows as $O(2^{2N})$, providing an advantage over the scaling $O(2^{4N})$ obtained for random Pauli measurements.

Crucially, shallow shadows require the computation of the inverse channel $\mathcal{M}^{-1}_t$. For global Clifford unitaries, a simple analytic formula is known~\cite{huang2020predicting}
	\begin{align}\label{eq:inversion_formula_infty}
  		\mathcal{M}_{\infty}^{-1}[\rho]&=(2^N+1)\rho-{\rm Tr}(\rho)\openone\,.
	\end{align} 
	However, no analogous result is available for finite $t>2$~\cite{bertoni2022shallow,arienzo2023closed,bu2024classical}. So far, numerical algorithms based on tensor networks (TNs)~\cite{silvi2019tensor} were developed to compute $\mathcal{M}_t^{-1}$ for 1D circuits, with a computational cost growing polynomially in $N$ and exponentially in $t$. At the same time, the problem remains intractable in higher dimensions or more general circuit connectivity. Overall, computing $\mathcal{M}_t^{-1}$ makes the post-processing step highly nontrivial ~\cite{bertoni2022shallow,akhtar2023scalable,arienzo2023closed}, limiting the scope and applicability of shallow shadows.

    \subsection{Approximate inversion formula}
    \label{sec:inversion_formula}

    In this section, we describe our proposed scheme to get around the challenging problem of computing and inverting $\mathcal{M}_t$, thus significantly expanding the power of shallow shadows. The main conceptual ingredient of our protocol is to simply replace the inverted measurement channel $\mathcal{M}_t^{-1}$ by its infinite-depth limit, $\mathcal{M}_\infty^{-1}$. That is, we introduce the approximate shadows 
    \begin{equation}
	\label{eq:approximate_shadow}\tilde\rho^{(r)}_t=\mathcal{M}_\infty^{-1}\left[(U_t^{(r)})^{\dagger}|b^{(r)}\rangle\langle b^{(r)}| U^{(r)}_t\right]\,.
 \end{equation}
This substitution gives rise to estimators which are not faithful for finite $t$, at which
\begin{equation}
    \mathbb{E}[\tilde\rho^{(r)}_t]\neq \rho\,.
\end{equation}
However, since we must recover $\mathcal{M}_{\infty}^{-1}[\mathcal{M}_t(\rho)]\simeq \rho$ for large $t$, we expect that the modified estimators become increasingly accurate as $t$ increases. The question is then what is the minimum depth $t^\ast$ at which the error is below some threshold $\delta$. 

Since the measurement channel depends on the unitary ensemble only through its second moment $\mathbb{E}[U_t\otimes U_t^{\ast}\otimes U_t\otimes U_t^{\ast}]$, one may guess that $t^\ast$ is the time taken for the second moment of the shallow-circuit ensemble to equilibrate to that of the global random Clifford ensemble. 
The latter is the same as the second moment of the Haar distribution due to the $2$-design property of the Clifford group~\cite{renes2004symmetric,ambainis2007quantum}. Thus, $t^\ast$ would coincide with the depth at which random circuits become a $2$-design~\cite{harrow2023approximate,mittal2023local}. In the following sections, we will clarify this intuition. We will show in particular that the time $t^\ast$ heavily depends on the quantity under consideration, the fidelity and purity estimation requiring a depth $O(\log N)$ and $O(N)$, respectively. 

Before leaving this section, it is important to comment on the fidelity estimator, which is obtained using the approximate inversion formula. Combining Eqs.~\eqref{eq:fidelity_estimator} and~\eqref{eq:approximate_shadow}, we obtain
\begin{equation}\label{eq:approximate_fidelity_estimator}
\tilde{\mathcal{F}}_t^{(e)}=\frac{1}{M}\sum_{r=1}^M \tilde{f}_t^{(r)}\,,
\end{equation}
where
\begin{equation}\label{eq:single_approximate_fidelity_estimator}
\tilde{f}^{(r)}_t=\langle{\psi} | \mathcal{M}_{\infty}^{-1}\left[(U_t^{(r)})^{\dagger}|b^{(r)}\rangle\langle b^{(r)}| U^{(r)}_t\right]|\psi\rangle\,.
\end{equation}
Interestingly, we can show that the expectation value of $\tilde{f}^{(r)}_t$ is closely related to a known function called linear cross entropy~\cite{arute2019quantum,neill2018blueprint}. To this end, we denote by $p_{\rho,U_t}(b)$ and $p_{\psi,U_t}(b)$, respectively, the probabilities to obtain the outcomes $b$ when the system is initialized in the states  $\rho$ and $\ket{\psi}$ and evolved by $U_t$. The average of $\tilde{f}^{(r)}_t$ can then be rewritten as
\begin{equation}
\mathbb{E}\left[\tilde{f}^{(r)}_t\right]=\mathbb{E}_{U_t\in \mathcal{U}}H_{\rm lin}[p_{\rho,U_t}, p_{\psi,U_t}]\,,
\end{equation}
where
\begin{equation}
    H_{\rm lin} [p,q]= 2^N\sum_b p(b)q(b)-1
\end{equation}
is the linear cross-entropy.
The linear cross entropy lies at the basis of the linear-cross-entropy benchmark (XEB). In this setting, one typically considers a trivial initial product state $\ket{0}$ and wants to certify the quantum circuit applied to it, which is assumed to be noisy~\cite{neill2018blueprint, boixo2018characterizing, arute2019quantum,liu2021benchmarking,wu2021strong,choi2023preparing,gao2024limitations,andersen2024thermalization}. Namely, XEB is typically used as a means to certify the fidelity of gates, rather than to perform quantum-state certification. In this work, instead, we assume that the gates are noiseless and Eq.~\eqref{eq:single_approximate_fidelity_estimator} is thus viewed as a tool to estimate the fidelity between the physical and target states.

We also mention an interesting connection with the protocol of filtered randomized benchmarking~\cite{helsen2022general}. Ref.~\cite{heinrich2022randomized}, in particular, used the inverse of the Haar-random measurement channel to define an estimator for the so-called filtered randomized benchmarking signal, which was shown to coincide with the linear-cross entropy.
    
Finally, we mention that $f^{(r)}_t$ is also related to another estimator considered in the literature. In particular, Ref.~\cite{choi2023preparing} introduced the normalized estimator
\begin{equation}
F_c=2\frac{\sum_zp_{\rho,U_t}(z)p_{\psi,U_t}(z)}{\sum_zp_{\rho,U_t}(z)^2}-1\,,
\end{equation}
see also \cite{shaw2023benchmarking,choi2023preparing,zhang2023superconducting}. $F_c$ was employed in the context of analog chaotic dynamics and it was demonstrated to estimate the fidelity at infinite effective temperature and short quench evolution times.

\section{Approximate fidelity estimator}
\label{sec:approximate_estimator_fidelity}

We now study the performance of the approximate fidelity estimator. To be concrete, we consider a quantum-certification protocol where the system is prepared in an unknown quantum state $\rho$. Given the target state $\ket{\psi}$, and any certification threshold $0<\varepsilon<1$, we want to determine whether $\mathcal{F}_\psi(\rho)$ is larger or smaller than $1-\varepsilon$.

\subsection{Gate complexity}
\label{sec:gate_complexity}

In order to establish the usefulness of the estimator $\tilde{\mathcal{F}}^{(e)}_t$, we first need to determine the depth at which the approximation error becomes small. To build intuition, we start from the ideal situation where the lab and target states coincide, $\rho=\ket{\psi}\bra{\psi}$, and ask what is the minimum depth $t_0^\ast(N,\delta)$ such that  
 \begin{equation}\label{eq:inequality_delta}
     |\mathbb{E}[\tilde{\mathcal{F}}_t^{(e)}]-1|\leq \delta\,,
 \end{equation}
for all $t\geq t_0^\ast(N,\delta)$, where $\delta>0$ is an arbitrary number that should be taken smaller than the certification threshold $\varepsilon$. This problem can be solved analytically when $\ket{\psi}$ is sufficiently simple, and in particular when $\ket{\psi}=\otimes_j \ket{\phi_j}$ is a product state. To this end, we use the $2$-design property of the Clifford group~\cite{zhu2016clifford}, stating that the averages over random Clifford  and Haar random gates are equal. Then, exploiting the invariance of the Haar measure with respect to single-qubit rotations, we derive in Appendix~\ref{sec:app_derivation_anticoncentration} the following result
	\begin{equation}\label{eq:final}
	\mathbb{E}\left[\tilde{\mathcal{F}}_t^{(e)}\right]=(2^N+1)2^NZ_t-1\,,
	\end{equation}
	where $Z_t=\mathbb{E}_{U_t}\left[|\braket{0|U_t|0}|^4\right]$, while $\ket{0}=\ket{0}^{\otimes N}$ and the average is over Haar random circuits of depth $t$.

	The average $Z_t$ can be analyzed employing recently-introduced mappings to statistical-mechanics partition functions~\cite{potter2022entanglement,fisher2022random}. In fact, $Z_t$ has been previously studied in the context of anticoncentration of random quantum circuits~\cite{turkeshi2024hilbert, dalzell2022random}. In particular, Ref.~\cite{dalzell2022random} provided rigorous lower and upper bounds on $Z_t$ which directly imply 
 \begin{equation}
 1\leq \mathbb{E}\left[\tilde{\mathcal{F}}_t^{(e)}\right]\leq 1+ 2e^{-a(t-T_N)}     
 \end{equation}
where the constant $a$ and $T_N$ (which depends on $N$) are determined by the circuit architecture. For 1D brickwork circuits, $a=\log(5/4)$ and $T_N=[\log N+\log( e-1)]/a + 1$, which finally implies $t_0^\ast(N,\delta)\leq g(N,\delta)$, where
	\begin{equation}\label{eq:t_star_bound}
		g(N,\delta) = \frac{\log (N/\delta)}{\log(5/4)} + \frac{\log[2(e-1)]}{\log (5/4)}+1\,,
	\end{equation}
yielding the anticipated logarithmic dependence on $N$. The same scaling $T_N\sim \log N$ (with a different prefactor) follows from analogous bounds on $Z_t$, which are proven rigorously for non-local quantum circuits and conjectured for more general architecture (including the $2D$ case) with minimal connectivity requirements~\cite{dalzell2022random}.
	
	In principle, one could extend the derivation of Ref.~\cite{turkeshi2024hilbert,dalzell2022random} to study more structured target states $\ket{\psi}$, but this is technically challenging. Here, we follow a different approach and provide a simple argument to establish a bound on $t^\ast(N,\delta)$ for \emph{typical} target states.
	
	Focusing first  on 1D architectures, we model a typical target state as $\ket{\psi}=W_\tau \ket{0}$, where $W_\tau$ is a 1D circuit of depth $\tau$.  The value of $\tau$ is arbitrary, but we assume that the circuit architecture is the same and that the gates are drawn from a Haar random distribution. This model is expected to capture the behavior of generic many-body quantum states away from fine-tuning. 
	
	For fixed $W_\tau$,  the computations leading to Eq.~\eqref{eq:final} yield
    \begin{equation}\mathbb{E}_{U_t}\left[\tilde{\mathcal{F}}_t^{(e)}\right]=(2^N+1)2^N\mathbb{E}\left[|\braket{0|U_tW_{\tau}|0}|^4\right]-1\,.
    \end{equation}
We extract the typical behavior by averaging over the circuits $W_\tau$ defining the target state and obtain
	\begin{equation}\label{eq:w_circuit_final}
	\mathbb{E}\left[\tilde{\mathcal{F}}_t^{(e)}\right]=(2^N+1)2^NZ_{t+\tau}-1\,.
	\end{equation}
	Denoting by $t_\tau^\ast(N,\delta)$ the time at which $|\mathbb{E}[\tilde{\mathcal{F}}_t^{(e)}]-1|\leq \delta$ for the target state $\ket{\psi}=W_\tau\ket{0}$, Eq.~\eqref{eq:w_circuit_final} immediately implies that  $t_\tau^\ast(N,\delta)\leq g(N,\delta)-\tau$, where $g$ is defined in Eq.~\eqref{eq:t_star_bound}.  That is, for typical states of constant complexity $\tau$ (modeled by Haar-random brickworks of depth $\tau$ acting on a product state), the bound on $t^\ast$ is smaller by an additive constant compared to that of product states. This is intuitive: we have related the accuracy of the estimator to the value of the anti-concentration parameter $Z_t$ of the evolved state---the estimator becomes accurate when the evolved state's wavefunction is spread uniformly over the computational basis~\cite{dalzell2022random}; but since typical product states are more concentrated 
    than typical time-evolved states, more ``scrambling'' is needed to achieve this condition when starting from product states. We can repeat this discussion for more general architectures, finding the same result. 

The previous arguments apply when the lab and target states coincide, but this is never realised in practice. In fact, $\rho$ will generally be mixed with $\mathcal{F}_\psi(\rho)<1$, making the analysis more complicated. In addition, we have always considered typical target states, but one could wonder whether our bounds apply to structured or fine-tuned target states. We tackle these questions numerically, and in Sec.~\ref{sec:numerical_results_fidelity} we provide evidence showing that our bounds remain indeed valid in these situations and that the depth $t^{\ast}$ displays a logarithmic scaling for all target states and circuit architectures.

\begin{figure*}[t!]
\centering
\includegraphics[trim={0 0.6cm 0 0},clip,width=\linewidth]{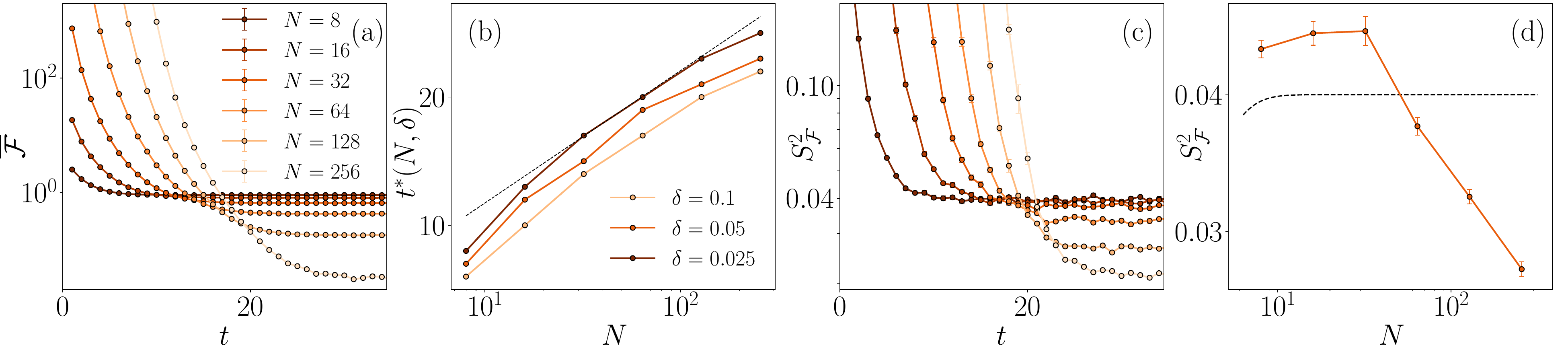}
\caption{Numerical simulation of the certification protocol for 1D local circuits. (a): Averaged fidelity estimator $\overline{\mathcal{F}}=\mathbb{E}[\tilde{\mathcal{F}}_t^{(e)}]$ as a function of the circuit depth, for increasing $N$. Here, the target state $\ket{\psi}$ is the GHZ state and $\rho=\mathcal{E}^{\otimes N}_p(\ket{\psi}\bra{\psi})$, where $\mathcal{E}_p$ is the depolarizing channel with $p=0.02$. Error bars associated with statistical fluctuations are small and not visible in the scales of the plots. (b): Depth $t^{\ast}(N,\delta)$ (defined in the main text) as a function of the system size $N$. The dashed line is the function in Eq.~\eqref{eq:t_star_bound}, shifted by a negative constant. (c): Variance of fidelity estimator $S^{2}_\mathcal{F}=\mathbb{E}[(\tilde{\mathcal{F}}_t^{(e)}-\mathcal{F}_\psi(\rho))^2]$ as a function of the depth. (d): The variance $S^{2}_\mathcal{F}$ estimated at the depth $t^\ast(N,\delta)$, for $\delta=0.05$. The dashed line is the exact variance for the global Clifford protocol when the lab state coincides with the target state.}
\label{fig:fig2}
\end{figure*}

We conclude this section with an important remark. Although our post-processing scheme gets around the inversion problem, evaluating the estimator ~\eqref{eq:single_approximate_fidelity_estimator} still requires one to compute amplitudes of the form $\langle{\psi} | (U_t^{(r)})^{\dagger}|b^{(r)}\rangle$. For generic states, the cost of this classical computation grows exponentially in $N$. However, the estimator can be computed efficiently (namely, in a time growing only polynomially in $N$) for different classes of target states. For 1D connectivity, this can be done, for instance, for TN states including Matrix Product States (MPS)~\cite{fannes1992finitely,perez2007matrix,cirac2017matrix_op} or tree tensor network states~\cite{silvi2019tensor}. In addition, in all circuit architectures, the amplitude can be computed efficiently for stabilizer states exploiting the classical simulability of Clifford circuits~\cite{gottesman1997stabilizer,gottesman1998theory}. This class includes the so-called Greenberger-Horne-Zeilinger (GHZ)~\cite{greenberger1989bell} and cluster state~\cite{briegel2001persisten} as special examples. We stress that the restriction to certain classes of target states is a common feature of efficient quantum certification protocols~\cite{huang2024certifying}.

\subsection{Sample complexity} 
\label{sec:sample_complexity}

In addition to the error due to the approximate inversion formula, one also needs to take into account the fluctuations of $\tilde{f}_t^{(r)}$ or, equivalently, of the estimator $\tilde{\mathcal{F}}^{(e)}_t$ (the variances of the two random variables are related by a factor $1/M$). In particular, the efficiency of the method depends on the possibility of bounding the variance of $\tilde{\mathcal{F}}^{(e)}_t$ by a constant independent of $N$. 

In standard classical shadow protocols, the estimator variances are bounded by the corresponding shadow norm squared~\eqref{eq:shadow_norm}. This object may be hard to study in general because it involves the third moment of the distribution over the random unitaries $U_t$. A more practical quantity is obtained by replacing the maximization in~\eqref{eq:shadow_norm} by the average over all states $\rho$, with a uniform measure (or any 1-design measure). This defines the \emph{locally-scrambled} shadow norm~\cite{bu2024classical,hu2023classical}, which is used as a proxy for $||O||_{\rm sh}$. For shallow shadows based on 1D circuits, the locally-scrambled shadow norm was studied in~ Refs.\cite{bertoni2022shallow,akhtar2023scalable,arienzo2023closed} and was proven to be $O(1)$ for circuits of depth $O(\log N)$~\cite{bertoni2022shallow}.

In Appendix~\ref{sec:appendix_variance}, we show that the variance of $\tilde{f}^{(r)}_t$ is bounded by an expression obtained replacing the exact inversion channel in~\eqref{eq:shadow_norm} by $\mathcal{M}_\infty^{-1}$. Following Refs.~\cite{bu2024classical,hu2023classical,akhtar2023scalable,bertoni2022shallow}, we then study its locally-scrambled version
\begin{align}
||O||^2_{\rm sh,LS}(t)=\frac{1}{2^N}\underset{U_t}{\mathbb{E}}\sum_{_{b \in\{0,1\}^N}}\left[ \left\langle b\left|U_t \mathcal{M}_\infty^{-1}(O) U_t^{\dagger}\right| b\right\rangle^2\right].\label{eq:local_shadow_norm}
\end{align}
This expression does not depend on $\rho$ and can be studied by repeating the arguments reported in Sec.~\ref{sec:inversion_formula}. 

Let us first consider the case where the target state is a product state, $\ket{\psi}=\otimes_j\ket{\phi_j}_j$. We can repeat the derivation leading to Eq.~\eqref{eq:final}, and by averaging over the random circuits $U_t$, we arrive at
\begin{equation}\label{eq:local_norm_value}
	||O||^2_{\rm sh,LS}(t)=(2^N+1)^2\left[Z_{t}-\frac{1}{2^{2N}}\right]\,.
\end{equation}
Therefore, $||O||_{\rm sh, LS}$ depends on the same term $Z_t$ appearing in the average in Eq.~\eqref{eq:final}. Now, recalling that $t_0^\ast(N,\delta)$ is defined as the minimum depth at which Eq.~\eqref{eq:inequality_delta} is satisfied, it follows from Eq.~\eqref{eq:final} that
\begin{equation}\label{eq:variance}
||O||^2_{\rm sh,LS}(t^\ast(N,\delta))\leq \frac{3}{2}(2+\delta)=O(1)\,.
\end{equation}
Following the previous section, we can finally invoke a typicality argument and provide a similar bound for generic states of the form $\ket{\psi}=W_\tau \ket{0}$ where $W_\tau$ is a random quantum circuit of depth $\tau$.

As mentioned, the locally-scrambled shallow norm is only a proxy for $||O||_{\rm sh}$. However, we expect that the $O(1)$ bound found above also holds for the actual shadow norm and thus for the variance of our estimator $\tilde{\mathcal{F}}^{(e)}_t$ whenever $t\geq t^\ast$. We substantiate this claim via extensive numerical evidence for different circuit architectures and target states below.

\subsection{Numerical results} 
\label{sec:numerical_results_fidelity}

\begin{figure*}[t]
\centering
\includegraphics[trim={0.5cm 0.5cm 0 0},clip,width=\linewidth]{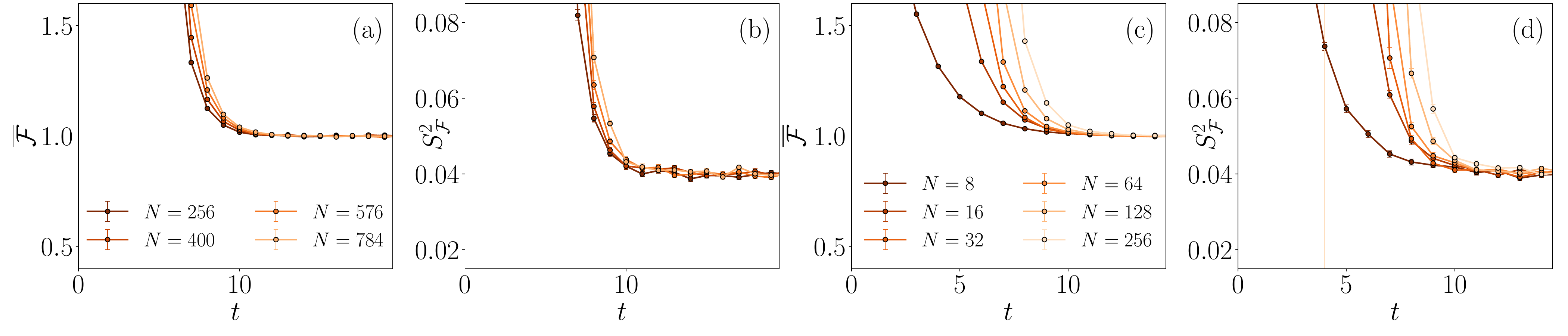}
\caption{Numerical simulation of the certification protocol for 2D local circuits [(a)-(b)] and non-local quantum circuits [(c)-(d)]. (a): Averaged fidelity estimator $\overline{\mathcal{F}}=\mathbb{E}[\tilde{\mathcal{F}}]$ as a function of the circuit depth, for increasing $N$. Here, the target state $\ket{\psi}$ is the $2D$ cluster state state and $\rho=\ket{\psi}\bra{\psi}$. The geometry is that of a square of side $L=\sqrt{N}$. (b): Variance of the fidelity estimator $S^{2}_\mathcal{F}=\mathbb{E}[(\tilde{\mathcal{F}}-\mathcal{F}(\rho))^2]$ as a function of the depth for the same state. (c): Averaged fidelity estimator $\overline{\mathcal{F}}=\mathbb{E}[\tilde{\mathcal{F}}]$ as a function of the circuit depth, for increasing $N$. Here, the target state $\ket{\psi}$ is the GHZ state and $\rho=\ket{\psi}\bra{\psi}$. (d): Variance of the fidelity estimator $S^{2}_\mathcal{F}=\mathbb{E}[(\tilde{\mathcal{F}}-\mathcal{F}(\rho))^2]$ as a function of circuit depth, for the same state. }
\label{fig:fig3}
\end{figure*}

In this section, we present numerical evidence supporting our claims and illustrate our scheme with explicit examples. We focus on two classes of states for which we can perform large-scale computations: 1D states and stabilizer states, both in one and two dimensions. For 1D states, we use an efficient method to compute the exact average over infinitely many measurement outcomes. This is based on the statistical mechanics mapping explained in Refs.~\cite{turkeshi2024hilbert, dalzell2022random} and numerical MPS computations. For stabilizer states, we use the classical simulability of Clifford circuits~\cite{gottesman1997stabilizer,gottesman1998theory,aaronson2004improvedsimulationof} to simulate the full quantum-certification protocol. We refer to Appendix~\ref{sec:appendix_numerics} for details on the numerical implementations, while we report representative examples of our data in Fig.~\ref{fig:fig2}--\ref{fig:fig3}: the data shown in these figures are obtained from Clifford simulations and checked independently against the MPS algorithm when possible.

We first consider the case where the lab state is a mixed state $\rho\neq \ket{\psi}\bra{\psi}$. In this case, we define analogously $t^\ast(N,\delta)$ as the minimum depth such that for $t\geq t^\ast$ one has $ \Delta F_{\psi,\rho}\leq \delta$, where
 \begin{equation}\label{eq:delta_F}
 \Delta F_{\psi,\rho}(t)=|\mathbb{E}[\tilde{\mathcal{F}}^{(e)}_t]-\mathcal{F}_{\psi}(\rho)|\,.
 \end{equation}
 In Fig.~\ref{fig:fig2}, the target state is the GHZ state, 
\begin{equation}\label{eq:ghz}
    \ket{\psi}=\frac{1}{\sqrt{2}}(\ket{0}^{\otimes N}+\ket{1}^{\otimes N}),
\end{equation}
while the lab state $\rho$ is taken to be $\rho=\mathcal{E}^{\otimes N}_p(\ket{\psi}\bra{\psi})$
where 
\begin{equation}
\mathcal{E}_p(\sigma)=(1-p)\sigma+p{\rm Tr}[\sigma]\frac{\openone}{2}\,,    
\end{equation}
is the depolarizing channel. Fig.~\ref{fig:fig2}(a) shows $\mathbb{E}[\tilde{\mathcal{F}}_t^{(e)}]$, for increasing $N$, while we report in Fig.~\ref{fig:fig2}(b) the value $t^\ast(N,\delta)$ at which $\Delta F_{\psi,\rho}(t)\leq \delta$, where $\Delta F_{\psi,\rho}(t)$ is defined in Eq.~\eqref{eq:delta_F}. We note that the GHZ state is a fine-tuned state, displaying long-range correlations and thus outside the scope of our analytical arguments. Yet, our numerical results confirm a logarithmic scaling of $t^\ast(N,\delta)$, for arbitrary small $\delta$. In fact, Fig.~\ref{fig:fig2}(b) displays the asymptotic scaling of $t^\ast$ predicted by Eq.~\eqref{eq:t_star_bound} (shifted by a negative constant), showing that it upper bounds $t^\ast(N,\delta)$ even though $\ket{\psi}$ is not a product state.

Going beyond the mean of the estimator, Fig.~\ref{fig:fig2}(c) and (d) show, respectively, the variance
\begin{equation}
    S^{2}_\mathcal{F}=\mathbb{E}\left[ \left(\tilde{\mathcal{F}}_t^{(e)} -\mathcal{F}_\psi(\rho)\right)^2 \right]\,,
\end{equation}
as a function of circuit depth $t$ and its value at the depth $t^\ast(N,\delta)$. The data clearly show that the variance quickly drops below an $N$-independent value. Interestingly, the latter is quantitatively close to that obtained in the protocol based on global Clifford unitaries, as we show in Appendix~\ref{sec:appendix_variance}. In the 1D case, we have further studied short-ranged correlated states, finding similar results, cf. Appendix~\ref{sec:appendix_numerics} for additional numerical data.

Next, we report data for circuit architectures beyond 1D in Fig.~\ref{fig:fig3}. 
Figs.~\ref{fig:fig3}(a,b) focus on local circuits on a 2D square lattice of side $L$. The target state is the cluster state, 
\begin{equation}
\ket{\psi}=\prod_{\langle i,j\rangle}CZ_{i,j} \ket{+}^{\otimes N}\,,    
\end{equation}
where the product is over nearest-neighbor qubits on the $2D$ lattice, while $\ket{+}=(\ket{0}+\ket{1})/\sqrt{2}$ and we assume $\rho=\ket{\psi}\bra{\psi}$. We are able to perform simulations for square circuits up to $N=L^2=784$ qubits. The depth $t^\ast(N,\delta)$ appears almost independent of the system size. We conjecture that this is a finite-size effect and that a logarithmic growth would be eventually observed at even larger values of $N$. Nevertheless, it is remarkable that an accuracy above $95\%$ can be obtained for $N\sim 800$ qubits by depth $t\lesssim 10$.

Finally, in Figs.~\ref{fig:fig3}(c,d) we study the protocol based on geometrically non-local quantum circuits, where the target state is the GHZ state. The plots clearly show that $t^\ast(N,\delta)$ increases at most logarithmically in $N$. As expected, we find that the non-local circuit requires a smaller depth in order to achieve the same accuracy for the GHZ state, compared to 1D local circuits. Again, this is intuitive, as the validity of our method relies on anticoncentration in quantum circuits, which happens faster on all-to-all connected circuit architectures~\cite{dalzell2022random}. For both the 2D and non-local circuits, the plots show that the variance approaches a finite constant at a depth $t^\ast(N,\delta)$ for arbitrary $\delta$. In summary, all cases that we have investigated fully support the claims presented in Sec.~\ref{sec:inversion_formula} about the efficiency of the protocol in terms of both gate complexity and sample complexity, even beyond the domain of validity of our analytical arguments.

In all our simulations, our estimator~\eqref{eq:fidelity_estimator} is built choosing $M=50$, while its average and variance are estimated sampling over $R\simeq 10000$ independent values. To be more precise, the total number of simulated experimental runs is $MR$. The outcomes are organized into $R$ subsets containing $M$ shadows each, which are used to compute an instance of the estimator~\eqref{eq:fidelity_estimator}. We stress that number of measurements performed in our simulations is of the same order as, or smaller than, the typical number of measurements performed in present-day trapped ion and superconducting qubit experiments~\cite{brydges2019probing,satzinger2021realizing,stricker2022experimental,zhu2022cross,hoke2023measurement,vitale2023estimation}, thus substantiating our claims on the practical feasibility of our scheme.

\section{Approximate purity estimator}
\label{sec:approximate_purity_estimator}

\begin{figure*}[t]
\centering
\includegraphics[trim={0.3cm 0.5cm 0 0},clip,width=\linewidth]{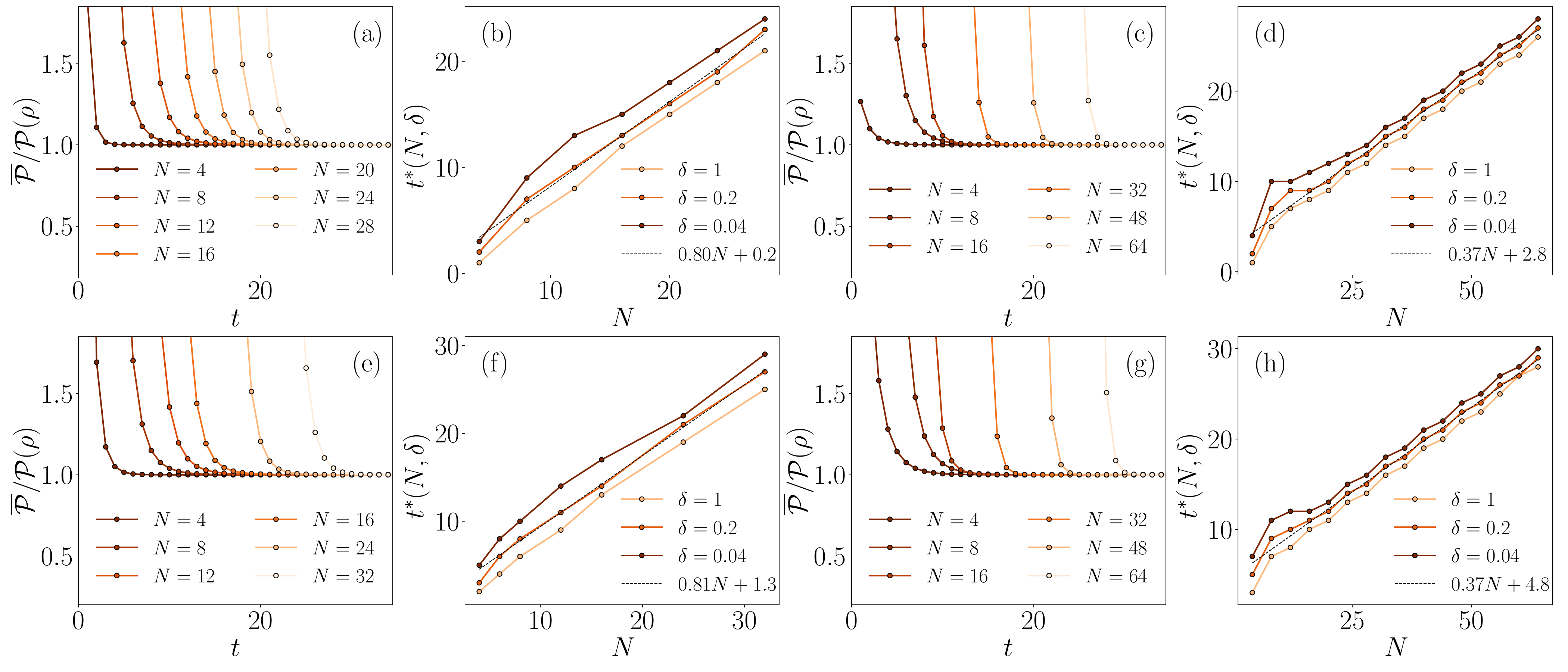}
\caption{Numerical results for the averaged purity estimator for 1D local [(a),(b),(e),(f)] and non-local circuits [(c),(d),(g),(h)]. In panels [(a)-(d)] the state $\rho$ is the GHZ state~\eqref{eq:ghz}, while in [(e)-(h)] it is the state given in Eqs.~\eqref{eq:product_1},~\eqref{eq:product_2} with $\mu=0.05$. Panels (a), (c), (e) and (g) show the ratio between the expected value of the purity estimator $\overline{\mathcal{P}}$ and the exact value of the purity $\mathcal{P}(\rho)$ as a function of circuit depth, for increasing $N$. Panels (b), (d), (f) and (h) show the depth $t^{\ast}(N,\delta)$ defined in~\eqref{eq:t_ast_purity}.}
\label{fig:purity_GHZ_mixed}
\end{figure*}

In this section, we study the performance of the approximate inversion formula for purity estimation. We consider, in particular, the estimator
\begin{equation}\label{eq:approximated_estimator_purity}
\tilde{\mathcal{P}}_t^{(e)}=\frac{1}{M(M-1)} \sum_{r \neq r^{\prime}} {\rm Tr}\left(\tilde\rho_{t}^{(r)} \tilde\rho_{t}^{\left(r^{\prime}\right)}\right)\,,
\end{equation}
where $\tilde{\rho}^{(r)}_t$ is defined in Eq.~\eqref{eq:approximate_shadow}. Being a quadratic function of $\tilde{\rho}_t$, this estimator is harder to treat analytically than the fidelity estimator, and its average cannot straightforwardly be related to the anti-concentration term $Z_t$ defined in Eq.~\eqref{eq:final}. Therefore, we study its behavior numerically, restricting for simplicity to the case of 1D and all-to-all circuits. As for the fidelity, we need to consider both the approximation error given by the approximate inversion formula and the statistical error associated with the random measurement outcomes. We present our results separately in the following.

The error due to the approximate inversion formula depends on the average 
\begin{equation}
\overline{\mathcal{P}}_t=\mathbb{E}\left[\tilde{\mathcal{P}}_t^{(e)}\right]\,,
\end{equation}
or equivalently on $\mathbb{E}[{\rm Tr}(\tilde{\rho}_t^{(r)}\tilde{\rho}_t^{(r')})]$. Since the purity $\mathcal{P}(\rho)$ is typically exponentially small in the system size $N$, it is natural to quantify the deviation of our estimator from the true value via the relative error. Accordingly, we define $t^\ast(N,\delta)$ as the minimum depth such that 
\begin{equation}\label{eq:t_ast_purity}
\left|\frac{\overline{\mathcal{P}}_t}{\mathcal{P}(\rho)}-1\right|\leq \delta\,,
\end{equation}
for all $t\geq t^\ast(N,\delta)$.

\begin{figure*}[t]
\centering
\includegraphics[trim={1.8cm 0.5cm 0 0},clip,width=0.8\linewidth]{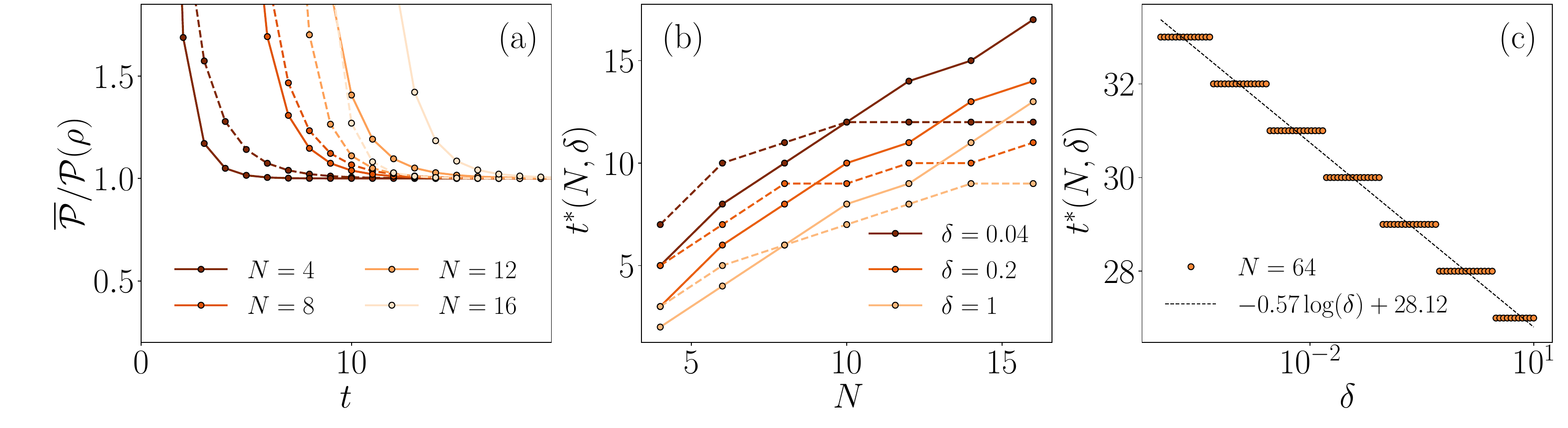}
\caption{Comparison of the 1D local architecture (solid lines) with the non-local one (dashed lines) for small system sizes $N$. Here the state $\rho$ is the product state given in Eqs.~\eqref{eq:product_1},~\eqref{eq:product_2} with $\mu=0.05$. Panel (c) shows the scaling of $t^{\ast}(N,\delta)$ with $\delta$ for $N=64$.}
\label{fig:purity_comparison}
\end{figure*}

We compute $\overline{\mathcal{P}}_t$ using numerically exact methods yielding the average over infinitely many shadows. As explained in Appendix~\ref{sec:details_purity}, these methods are based on a standard replica formalism~\cite{potter2022entanglement,fisher2022random}, mapping the problem onto a non-unitary dynamics over $N$ qubits. In the 1D geometry, the latter dynamics can be computed efficiently using TN algorithms. In the non-local geometry, there is an emergent permutation symmetry, allowing us to perform computations up to very large system sizes, cf. Appendix~\ref{sec:appendix_nonlocal}. In the following, we report the data obtained using these methods. We also mention that, when possible, we have checked these results against a direct simulation of the full purity-estimation protocol using the stabilizer formalism (namely, collecting a finite number of shadows and averaging over them). However, since the variance of the purity estimator grows exponentially in the number of qubits $N$, cf.~\eqref{eq:variance_purity_global}, this approach is limited to rather small system sizes. 

We study two examples. First, we consider the GHZ state~\eqref{eq:ghz}, for which the purity is one, and the product state
\begin{equation}\label{eq:product_1}
\rho = \bigotimes_{j=1}^N \rho_j\,,
\end{equation}
where 
\begin{equation}\label{eq:product_2}
\rho_j = \cos^2(\mu)|0\rangle\langle 0| + \sin^2(\mu)|1\rangle \langle 1|\,,
\end{equation}
for which the purity is
\begin{equation}\label{eq:purity_true_value}
\mathcal{P}(\rho) = (\cos^4(\mu)+\sin^4(\mu))^N\,.    
\end{equation}

In Fig.~\ref{fig:purity_GHZ_mixed} we show our results for the GHZ and product state~\eqref{eq:product_1}, respectively in panels (a)-(d) and (e)-(h). In each row, the first two panels correspond to the 1D local architecture, while the last two to the non-local circuits. Panels (a) and (c), (e) and (g) show the ratio between the expectation value of the purity estimator and the true value as a function of the
circuit depth for increasing values of $N$. Interestingly, we find that for both circuit architectures and all considered states, $t^{\ast}(N,\delta)$ is linear in $N$, with a coefficient that is fit to $0.81\pm 0.02$ for 1D local circuits and $0.37 \pm 0.01$ for non-local ones, for both the GHZ state and the product state with $\mu=0.05$ (for increasing - but small - values of $\mu$ we find that the linear coefficient increases slightly). This implies that the non-local architecture performs better for large values of $N$, as expected. For small $N$, however, we show in Fig.~\ref{fig:purity_comparison} that this is not necessarily true. We stress that in all cases the leading term in $N$ of $t^{\ast}(N,\delta)$, which is proportional to $N$, appears to be independent of $\delta$, at least for the values of $N$ that we could simulate. More quantitatively, Fig.~\ref{fig:purity_comparison}(c) shows the dependence of $t^\ast(\delta,N)$ on $\delta$ (for fixed $N$). The plot is consistent with the scaling $t^\ast(N,\delta)\propto \log(1/\delta)$.

\begin{figure*}[t]
\centering
\includegraphics[trim={1.5cm 0.5cm 0 0},clip,width=\linewidth]{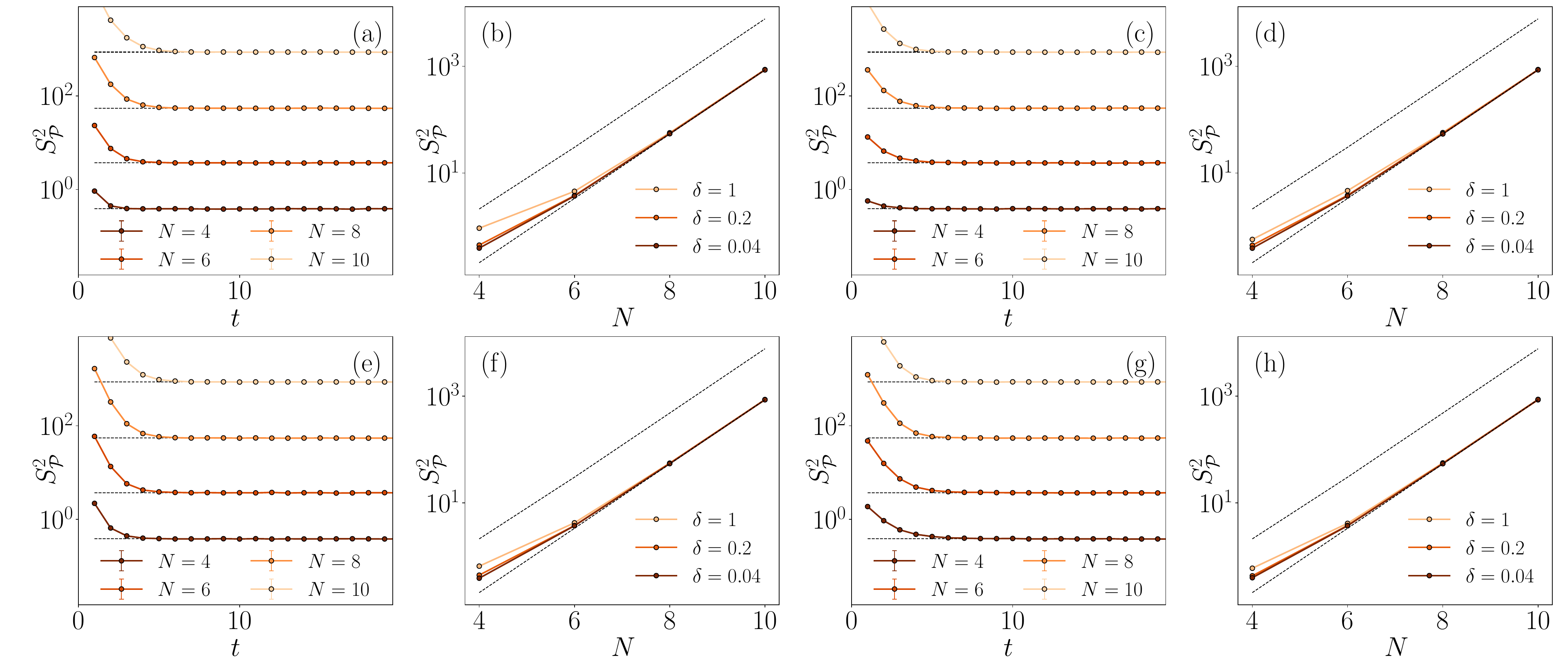}
\caption{Numerical computation of the variance of the purity estimator for 1D local circuits [(a),(b),(e),(f)] and non-local ones [(c),(d),(g),(h)]. In panels [(a)-(d)] the state $\rho$ is the GHZ state~\eqref{eq:ghz}, while in [(e)-(h)] it is the state given in Eqs.~\eqref{eq:product_1},~\eqref{eq:product_2} with $\mu=0.05$. Panels (a), (c), (e) and (g) show the sample variance as a function of circuit depth for increasing $N$. The dashed lines correspond to the $t \to \infty$ value in Eq.~\eqref{eq:variance_purity_exact}. Panels (b), (d), (f) and (h) display the variance at depth $t^{\ast}(N,\delta)$ for increasing $N$, for different values of $\delta$. The lower dashed line represents the scaling in Eq.~\eqref{eq:purity_scaling}, while the upper one shows the upper bound in Eq.~\eqref{eq:variance_purity_global}.}
\label{fig:purity_variance}
\end{figure*}

Next, we study the variance of the purity estimator
\begin{equation}
S_{\mathcal{P}}^2 = \mathbb{E}[(\tilde{\mathcal{P}}_t^{(e)} - \mathbb{E}[\tilde{\mathcal{P}}_t^{(e)}])^2]\,,
\end{equation}
which determines the sample complexity of the protocol. For both the GHZ and the product state~\eqref{eq:product_1}, we simulate the sampling of shadows using efficient simulations of Clifford circuits and report our results in Fig.~\ref{fig:purity_variance}. For each row, the first two panels correspond to the 1D local architecture, while the last two to the non-local circuits. Panels (a), (c), (e) and (g) show $S_{\mathcal{P}}^2$ as a function of circuit depth, for increasing values of $N$, while in (b), (d), (f) and (h) we report the same variance evaluated at $t^{\ast}(N,\delta)$, as a function of $N$. The plots show that in both cases, the variance approaches the global-Clifford value $\mathrm{Var}[\tilde{\mathcal{P}}_{\infty}^{(e)}]$ at the depth $t^{\ast}(N,\delta)$, for arbitrary $\delta$. We can compute this variance analytically, improving the bound~\eqref{eq:variance_purity_global} derived for general observables in Ref.~\cite{huang2020predicting}, cf. Appendix~\ref{sec:appendix_purity_variance}. In the large-$N$ limit, it simplifies to
\begin{equation}\label{eq:purity_scaling}
    \mathrm{Var}[\tilde{\mathcal{P}}_{\infty}^{(e)}] \sim \frac{2(2^{2N})}{M(M-1)}\,.
\end{equation}
As already anticipated, the variance grows exponentially in the system size, but it is still a quadratic improvement over the $O(2^{4N})$ scaling of the random Pauli measurements protocol. This makes our protocol better suited to estimate subsystem purities, which have been used to detect pure state entanglement \cite{brydges2019probing, satzinger2021realizing, stricker2022experimental, zhu2022cross, hoke2023measurement} for small system sizes.

It is also worth mentioning that our approximate method could even be useful to measure entanglement-related quantities for large system sizes, under some additional assumptions. For instance, Ref.~\cite{vermersch2023many} introduced a protocol to measure the bipartite entanglement of pure states in the absence of long-range correlations, using only polynomially many measurements. In that protocol, the global purity is reconstructed starting from the purities of finite subregions. Ref.~\cite{vermersch2023many} proposed to measure such purities using local Pauli measurements, but one could instead use our approximate scheme to make the protocol even more efficient (the number of measurements would remain polynomial, but with an improved scaling).

In all the Clifford simulations, the estimator~\eqref{eq:estimator_purity} is constructed choosing $M$ = 50, while its average and variance are estimated by sampling over $R \approx$ 90000 independent values. This specific value of $R$ was chosen to guarantee a relative error on the purity estimator smaller than $10\%$ for all system sizes up to $N=10$.

\section{Application to Pauli observables}
	
	\begin{figure*}[t]
		\centering
		\includegraphics[trim={0.5cm 0.5cm 0 0},clip,width=\linewidth]{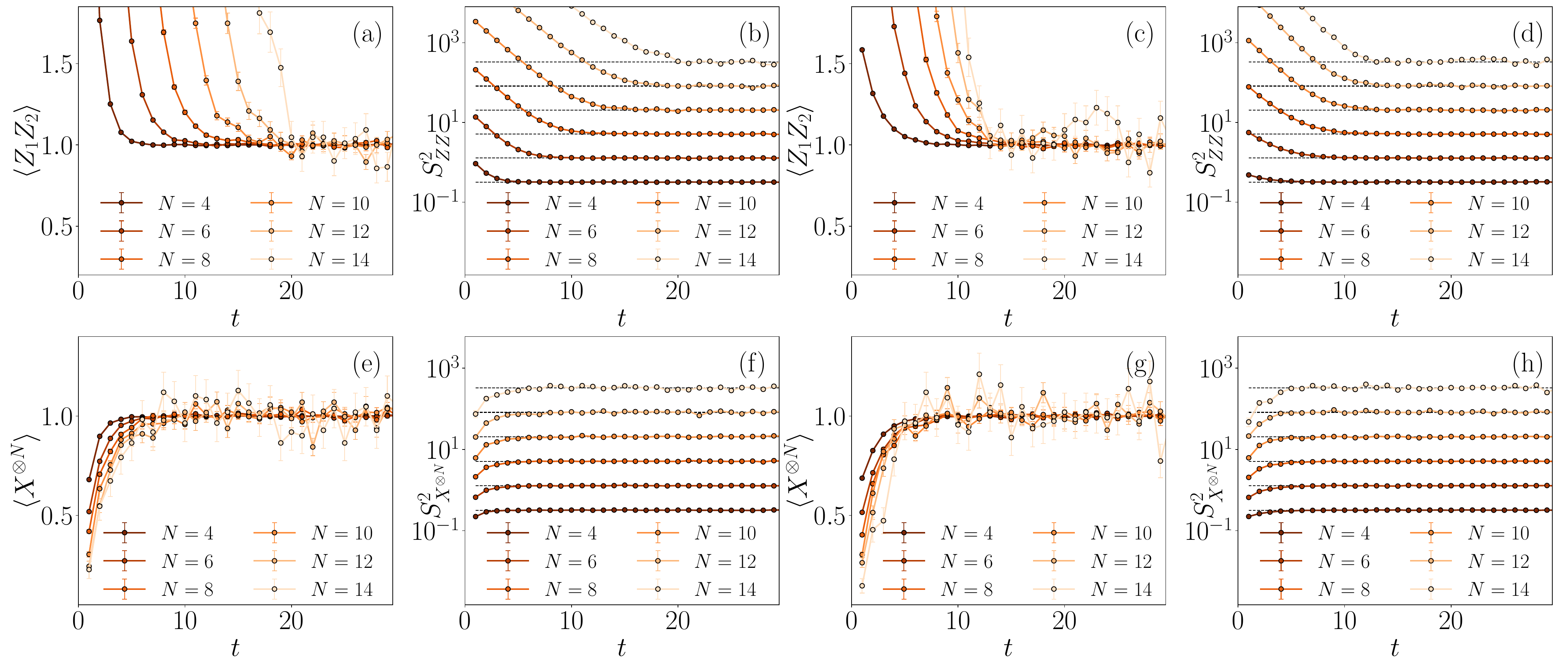}
		\caption{Numerical computation of the average of the estimator for expectation values of Pauli operators, for 1D local circuits [(a),(b),(e),(f)] and non-local ones [(c),(d),(g),(h)]. The initial state is the GHZ state~\eqref{eq:ghz}. Panels (a), (c), (e) and (g) show the average value as a function of circuit depth for increasing $N$. Panels (b), (d), (f) and (h) show the variance for the same depths and sizes. The dashed lines correspond to the $t \to \infty$ value in Eq.~\eqref{eq:variance_expval}.}
		\label{fig:expval_ghz}
	\end{figure*}

	Finally, we briefly discuss how the approximate inversion protocol performs in computing expectation values of Pauli observables. For a given Pauli $P$, we construct the approximate estimator 
	\begin{equation}
		\label{eq:pauli_estimator}
		\tilde{P}_t^{(e)} = \frac{1}{M}\sum_{r=1}^M \mathrm{Tr}\left(P \tilde{\rho}_t^{(r)} \right),
	\end{equation}
	with $\tilde{\rho}_t^{(r)}$ defined in Eq.~\eqref{eq:approximate_shadow}. We consider the case where the lab state is the GHZ state, and study the behaviour of the protocol numerically using Clifford-circuit simulations.  In the standard shadow protocols, it is known that qualitative differences in the scaling of the variance arise depending on the locality of the Pauli operators. Therefore, we study to examples of local and non-local observables, namely $Z_1 Z_2$ and $X^{\otimes N}$, respectively. Our results are shown in Fig.~\ref{fig:expval_ghz}. Once again, the first two panels of each row correspond to the 1D local architecture, while the last two to non-local circuits. Panels (a), (c), (e) and (g) show the behaviour of the expectation value of the approximate estimator as a function of circuit depth $t$ and for increasing number of qubits $N$, while panels (b), (d), (f) and (h) show its variance. 
	
	In all cases, for large enough depths we  find that the variance approaches the global-Clifford value of
	\begin{equation}
		\label{eq:variance_expval}
		\mathrm{Var}[\tilde{P}_{\infty}^{(e)}] = \frac{2^N+1-\langle P \rangle_{\psi}^2}{M}\,.
	\end{equation}
	The $t \to \infty$ value of the variance was computed following the same steps as in Appendix~\ref{sec:appendix_variance}. Since the variance is exponentially large in $N$, we are limited to rather small system sizes. Therefore, we are not always able to carry out a precise analysis of the scaling of the circuit depth $t^{\ast}(N, \delta)$ at which the estimator becomes unbiased. Yet, our data show clearly that  $t^{\ast}(N, \delta)$ is bounded by $O(N)$, which is consistent with the expected depth required for the random circuits to become a $2$-design.
	
	In fact, our data are consistent with a linear growth in $N$ of $t^{\ast}(N, \delta)$ in the case of $Z_1Z_2$. We stress that for local observables the protocol involving local Pauli measurements is exponentially better than that of global random Cliffords, so we do not expect our inverse approximate method to be useful in this case. Conversely, for the non-local observable $X^{\otimes N}$, the asymptotic scaling of $t^\ast(N,\delta)$ is not clear from our limited system sizes. Yet, our data appear to be consistent with a sub-linear growth of $t^\ast(N,\delta)$ with $N$. In this case, the variance associated with the global Clifford protocol grows as $O(2^N)$, which is an improvement over the $O(3^N)$ scaling of local Pauli measurements, so our method could be practically useful in this case.
	
	The estimator~\eqref{eq:pauli_estimator} is constructed with $M$ = 50, while its average and variance are estimated sampling over $R =$ 36000 independent values, such that the relative error on the Pauli estimator is smaller than $10\%$ for all system sizes up to $N=14$.

\section{Outlook} 
\label{sec:outlook}

We have introduced a simple approximate inversion formula for shallow shadows, and studied its performance for the global fidelity and purity estimation. Through analytic arguments and numerical computations, we have shown that the fidelity and purity estimators become accurate at depths scaling as $O(\log N)$ and $O(N)$, respectively. In addition, in both cases, we have shown that the estimator variances at those depths display the same scaling with $N$ as those obtained via global random unitaries. We have performed extensive numerical simulations showing that a modest amount of resources suffices in practice in many cases of interest, making the protocol relevant for current quantum computing platforms. We also mention that our approximate scheme may be useful beyond the applications illustrated in this work. For instance, it could be immediately combined with a recent scheme for efficient purity estimation in particular classes of many-body states~\cite{vermersch2023many}  to reduce the sample complexity further. 

Our work naturally raises the question of whether improved approximate inversion formulas can be found, possibly depending on the circuit architecture or the quantity to be estimated. Similarly to the one studied in our paper, two sensible requirements are that the approximate inversion formula can be easily computed in practice and that the corresponding estimators become accurate at modest circuit depth. It would be very interesting if such approximate formulas could be derived.

Finally, in order to further extend the applicability of our protocol, a natural direction would be to study its compatibility with available error-mitigation techniques~\cite{temme2017error,mcArdle2019error,koczor2021exponential,czarnik2021error,strikis2021learning,piveteau2022quasiprobability,guo2022quantum,van2023probabilistic,filippov2023scalable, mangini2024tensor}. In fact, we expect that relatively simple error mitigation strategies can be incorporated to take into account the effect of noise in the shallow quantum circuits used to perform the classical shadow measurements. We leave this interesting topic for future work.\\

\noindent \emph{Note added}. While finalising this draft, Refs.~\cite{schuster2024random} and ~\cite{laRacuente2024approximate} appeared on the arXiv, which rigorously show that quantum circuits form approximate unitary designs over $N$ qubits in $\log(N)$-depth. An immediate consequence of this result is that the approximate estimator for the fidelity studied in Sec.~\ref{sec:approximate_estimator_fidelity} becomes accurate in depth $O(\log N)$, which is consistent with our findings. The approximation measure used in Ref.~\cite{schuster2024random,laRacuente2024approximate} (relative error) does not directly constrain the purity estimator, so the results are also consistent with our findings in Sec.~\ref{sec:approximate_purity_estimator}. 

\begin{acknowledgements}
We are very grateful to Benoit Vermersch for valuable feedback on the manuscript. M.I. acknowledges helpful discussions with Robert Huang, Nick Hunter-Jones, and Vedika Khemani. 
X.T. acknowledges DFG under Germany's Excellence Strategy – Cluster of Excellence Matter and Light for Quantum Computing (ML4Q) EXC 2004/1 – 390534769, and DFG Collaborative Research Center (CRC) 183 Project No. 277101999 - project B01.
The research of R.C. and E.E. is partially funded by INFN (project “QUANTUM”). 
E.E. also acknowledges financial support from the PRIN Project 2022H77XB7 “Hybrid algorithms for quantum simulators”.
This work was co-funded by the European Union (ERC, QUANTHEM, 101114881). Views and opinions expressed are however those of the author(s) only and do not necessarily reflect those of the European Union or the European Research Council Executive Agency. Neither the European Union nor the granting authority can be held responsible for them.\\ 

\textbf{Data Availability Statement.---}
The source code and the data have been deposited in the Zenodo public folder \cite{turkeshi_2025_15118326}. 
\end{acknowledgements}

	
\appendix	
\begin{widetext}
	
\section{Details on the approximate fidelity estimator}
\label{sec:appendix_mapping_stat_mech}

This Appendix provides additional details on the derivations presented in Ref.~\ref{sec:approximate_estimator_fidelity}.

\subsection{Derivation of Eq.~\eqref{eq:final}}
\label{sec:app_derivation_anticoncentration}
First, we note that the average of the fidelity estimator~\eqref{eq:approximate_fidelity_estimator} and the variable $\tilde{f}^{(r)}_t$ in Eq.~\eqref{eq:single_approximate_fidelity_estimator} coincide and we may therefore consider the latter. Using the definition of the inverse channel ~\eqref{eq:inversion_formula_infty}, we can rewrite
\begin{equation}
 \tilde{f}^{(r)}_t=\langle{\psi} | (2^N+1)\left[(U_t^{(r)})^{\dagger}|b^{(r)}\rangle\langle b^{(r)}| U^{(r)}_t\right]|\psi\rangle - 1\,,
\end{equation}
where $b^{(r)} = 0, ..., 2^N-1$ is a bitstring sampled with probability $p_{b,U}^{(r)} = \langle{b^{(r)}}|U_t^{(r)}\rho (U_t^{(r)})^{\dagger}|b^{(r)}\rangle$ and $U_t^{(r)}$ a depth-$t$ random circuit made of two-qubit Clifford gates. We consider the case in which the lab state and the target state coincide, $\rho = |\psi\rangle\langle\psi|$, and $\ket{\psi}$ is a product state, $\ket{\psi}=\otimes_j \ket{\phi}_j$. 

The expectation value reads
\begin{align}
    \mathbb{E}\left[\tilde{f}^{(r)}_t\right]=(2^N+1)\sum_{b^{(r)} = 0}^{2^N-1} \int d\mu(U_t^{(r)}) |\langle{b^{(r)}}|U_t^{(r)}|\psi\rangle|^2 p_{b,U}^{(r)}- 1\,,
\end{align}
where we replaced the sum over $2$-qubit Clifford gates by the average over Haar-random gates due to the $2$-design property of the Clifford unitaries~\cite{zhu2016clifford}. Since $\ket{\psi}$ and $\ket{b^{(r)}}$ are product states, we can express them as products of single-qubit unitaries acting on $\ket{0} = \otimes_j\ket{0}_j$. By invoking the invariance of the Haar measure for left and right multiplication, we then obtain the expression
\begin{align}
    \mathbb{E}\left[\tilde{\mathcal{F}}_t^{(e)}\right]&= (2^N+1)2^N \int d\mu(U_t) |\braket{0|U_t|0}|^4 - 1 = (2^N+1)2^NZ_t-1\,,
\end{align}
where we introduced $Z_t=\mathbb{E}_{U_t}\left[|\braket{0|U_t|0}|^4\right]$. 

The term $Z_t$ was studied extensively in Ref.~\cite{dalzell2022random}. There, a notion of collision probability $\sum_b |\braket{b|U_t|0}|^4$ was introduced, expressing the probability that two independent measurements of two copies of the state $U_t\ket{0}$ yield the same outcome. Averaging over the random circuit instances $U_t$ we obtain $Z_t$ (up to a $2^N$ coefficient), which has a minimum when the probability distribution $p_{b,U} = |\braket{b|U_t|0}|^2$ is uniform.
This property was used to define the notion of anti-concentration: a random circuit $U_t$ is said to anti-concentrate if $Z_t$ differs at most by a constant value from its minimum.

Crucially, Ref.~\cite{dalzell2022random} also provided upper and lower bounds on the collision probability for different circuit architectures. For a general depth-$t$ random circuit with two-qubit gates, the following bounds were found~\cite{dalzell2022random}
\begin{equation}\label{eq:appendix_bounds}
    2 \leq 2^N(2^N+1) Z_t \leq 2 + 2e^{-a(t-T_N)}\,,
\end{equation}
where $a$ and $T_N$ depend on the circuit architecture.
Thus, the inequality $\mathbb{E}\left[\tilde{\mathcal{F}}_t^{(e)}\right] - 1 \leq \delta$ (for $\delta>0$), is equivalent to
\begin{equation}\label{eq:inequality}
   \mathbb{E}\left[\tilde{\mathcal{F}}_t^{(e)}\right] - 1 = (2^N+1)2^N Z_t - 2 \leq 2 e^{-a(t-T_N)} \leq \delta\,.
\end{equation}
This inequality immediately implies that the error on the fidelity becomes smaller than $\delta$ at a time $t_0^{\ast}(N,\delta) \leq g(N,\delta)$ with
\begin{equation}\label{eq:appendix_tstar}
    g(N,\delta) = T_N + a^{-1}\log(2/\delta).
\end{equation}
In the case of 1D brickwork circuits, $a = \log(5/4)$ and $T_N = [\log(N)+\log(e-1)]/a + 1$~\cite{dalzell2022random}, which finally implies Eq.~\eqref{eq:t_star_bound}.

\subsection{Variance of the approximate estimator}
\label{sec:appendix_variance}

We start from the definition of the locally-scrambled shadow norm for the operator $O$, which is obtained by replacing the maximization in Eq.~\eqref{eq:shadow_norm} with the average over the ensemble $\{V\rho V^{\dagger}| V=\otimes_j V_j \in \mathrm{U}(2)^{\otimes N} \}$
\begin{align}
    \notag
    ||O||^2_{\rm sh,LS} & = \underset{V,U_t}{\mathbb{E}}\sum_{_{b \in\{0,1\}^N}}\left[\left\langle b\left|U_t V\rho V^{\dagger} U_t^{\dagger}\right| b\right\rangle \left\langle b\left|U_t \mathcal{M}_\infty^{-1}(O) U_t^{\dagger}\right| b\right\rangle^2\right] \\
    & = \frac{1}{2^N} \underset{U_t}{\mathbb{E}}\sum_{_{b \in\{0,1\}^N}}\left[ \left\langle b\left|U_t \mathcal{M}_\infty^{-1}(O) U_t^{\dagger}\right| b\right\rangle^2\right],
\end{align}
where we used the fact that $\mathbb{E}_V \left[V\rho V^{\dagger}\right] = \openone/2^N$. From here (Eq.~\eqref{eq:local_shadow_norm}) we can replace $\mathcal{M}_{\infty}^{-1}(O) = (2^N+1)O$ and evaluate
\begin{align}
    \notag
    ||O||^2_{\rm sh,LS} & = \frac{(2^N+1)^2}{2^N}\int d\mu(U_t) \sum_{b=0}^{2^N-1}\left\langle b\left|U_t\left(|\psi\rangle\langle\psi| - \frac{\openone}{2^N}\right)U_t^{\dagger}\right|b\right\rangle^2 \\
    \notag
    & = (2^N+1)^2 \int d\mu(U_t) \left[|\braket{0|U_t|\psi}|^4 - \frac{2}{2^N} |\braket{0|U_t|\psi}|^2 + \frac{1}{2^{2N}}  \right] \\
    & = (2^N+1)^2 \left( \int d\mu(U_t) |\braket{0|U_t|\psi}|^4 - \frac{1}{2^{2N}} \right) = (2^N+1)^2 \left(Z_t - \frac{1}{2^{2N}} \right),
\end{align}
where the last equality (Eq.~\eqref{eq:local_norm_value}) holds if $\ket{\psi}$ is a product state. As before, after a depth $t^{\ast}(\delta, N)$ we have $2 \leq (2^N+1)2^N Z_{t^{\ast}} \leq 2 + \delta$ and thus $||O||^2_{\rm sh,LS} \leq (1+2^{-N})(2+\delta) \leq 3(2+\delta)/2$.

We can also compute explicitly the variance of $\tilde{\mathcal{F}}_t^{(e)}$ (Eq.~\eqref{eq:approximate_fidelity_estimator}) at late times. We start from
\begin{equation}
\begin{split}
    \mathbb{E}\left[(\tilde{\mathcal{F}}_t^{(e)})^2\right] & = \mathbb{E} \left[\frac{(2^N+1)^2}{M^2}\sum_{r,s=1}^{M}|\braket{b^{(r)}|U_t^{(r)}|\psi}|^2|\braket{b^{(s)}|U_t^{(s)}|\psi}|^2 - 2\frac{(2^N+1)}{M}\sum_{r=1}^{M} |\braket{b^{(r)}|U_t^{(r)}|\psi}|^2 + 1 \right]\,.
\end{split}
\end{equation}
The expectation value of the second term when $t \to \infty$ is just $-2(\mathbb{E}[\tilde{\mathcal{F}}_t^{(e)}]+1) = -4$. The first term can be rewritten as
\begin{equation}
    \frac{(2^N+1)^2}{M^2}\sum_{r}\mathbb{E}\left[|\braket{b^{(r)}|U_t^{(r)}|\psi}|^4\right] + \frac{(2^N+1)^2}{M^2}\sum_{r\neq s}\mathbb{E}\left[|\braket{b^{(r)}|U_t^{(r)}|\psi}|^2\right] \mathbb{E}\left[|\braket{b^{(s)}|U_t^{(s)}|\psi}|^2\right],
\end{equation}
where in turn the second term is aymptotically
\begin{equation}
    \frac{1}{M^2}\sum_{r\neq s}(\mathbb{E}[\tilde{f}^{(r)}_t]+1)^2 = \frac{4(M-1)}{M},
\end{equation}
and the first term
\begin{equation}
    \frac{(2^N+1)^2}{M^2} \sum_{r=1}^{M} \int d\mu(U_t) \sum_{b^{(r)}=0}^{2^N-1} |\braket{b^{(r)}|U_t^{(r)}|\psi}|^6 = \frac{2^N(2^N+1)^2}{M}\int d\mu(U_t) \mathrm{Tr}(|0\rangle\langle 0|^{\otimes 3} U_t^{\otimes 3} |\psi\rangle\langle\psi|^{\otimes 3} (U_t^{\dagger})^{\otimes 3})
\end{equation}
can be computed at late times by replacing the $d\mu(U_t)$ with the Haar measure over the full $U(2^N)$ unitary group. In this case, we have
\begin{equation}\label{eq:haar3}
    \int d\mu_H(U) U^{\otimes 3} |\psi\rangle\langle\psi|^{\otimes 3} (U^{\dagger})^{\otimes 3}) = \frac{1}{2^N(2^N+1)(2^N+2)} \sum_{\sigma \in S_3} P_{\sigma},
\end{equation}
where $P_{\sigma}$ is the permutation operator implementing the permutation $\sigma$ on the three replicas of the system. Thus the above equation yields $\frac{1}{M}\frac{6(2^N+1)}{(2^N+2)}$. Putting everything together, we get
\begin{equation}
    \mathrm{Var}\left[\tilde{\mathcal{F}}_t^{(e)}\right] =  \mathbb{E}\left[(\tilde{\mathcal{F}}_t^{(e)})^2\right] - 1 = \frac{1}{M}\frac{6(2^N+1)}{(2^N+2)} + \frac{4(M-1)}{M} - 4 = \frac{1}{M}\frac{2(2^N-1)}{2^N+2},
\end{equation}
which is a factor $1/M$ smaller than the variance of $\tilde{f}^{(r)}_t$.


\subsection{Numerical methods and additional results}
\label{sec:appendix_numerics}

\begin{figure*}[t]
\centering
\includegraphics[trim={1cm 0.6cm 0 0},clip,width=0.6\linewidth]{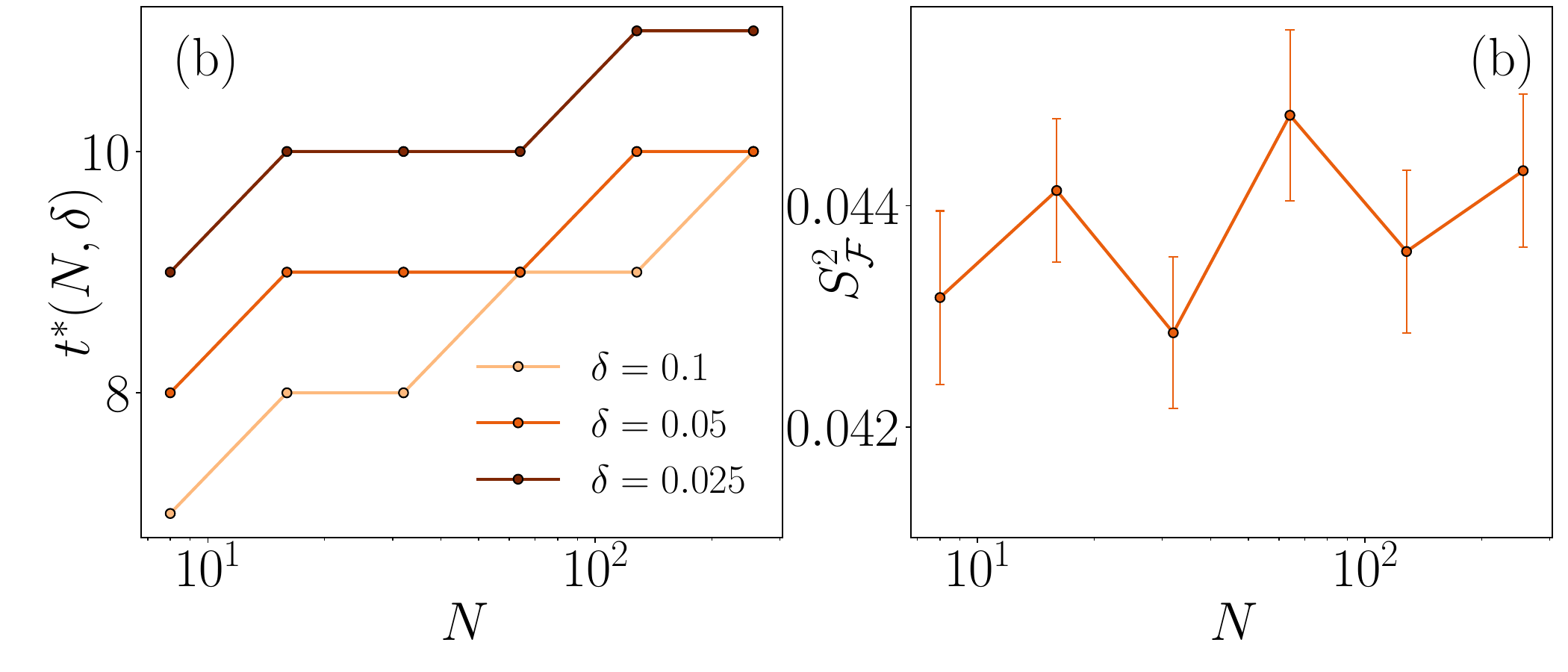}
\caption{Numerical simulation of the certification protocol for 1D non-local circuits, where both the lab state and the target state are the GHZ state. (a): Depth $t^{\ast}(N,\delta)$ after which $\Delta F_{\psi,\psi}\leq \delta$. (b): The variance $S^{2}_\mathcal{F}$ estimated at the depth $t^\ast(N,\delta)$, for $\delta=0.05$.}
\label{fig:C1}
\end{figure*}

\begin{figure*}[t]
\includegraphics[trim={0.6cm 0.6cm 0.5cm 0},clip,width=\linewidth]{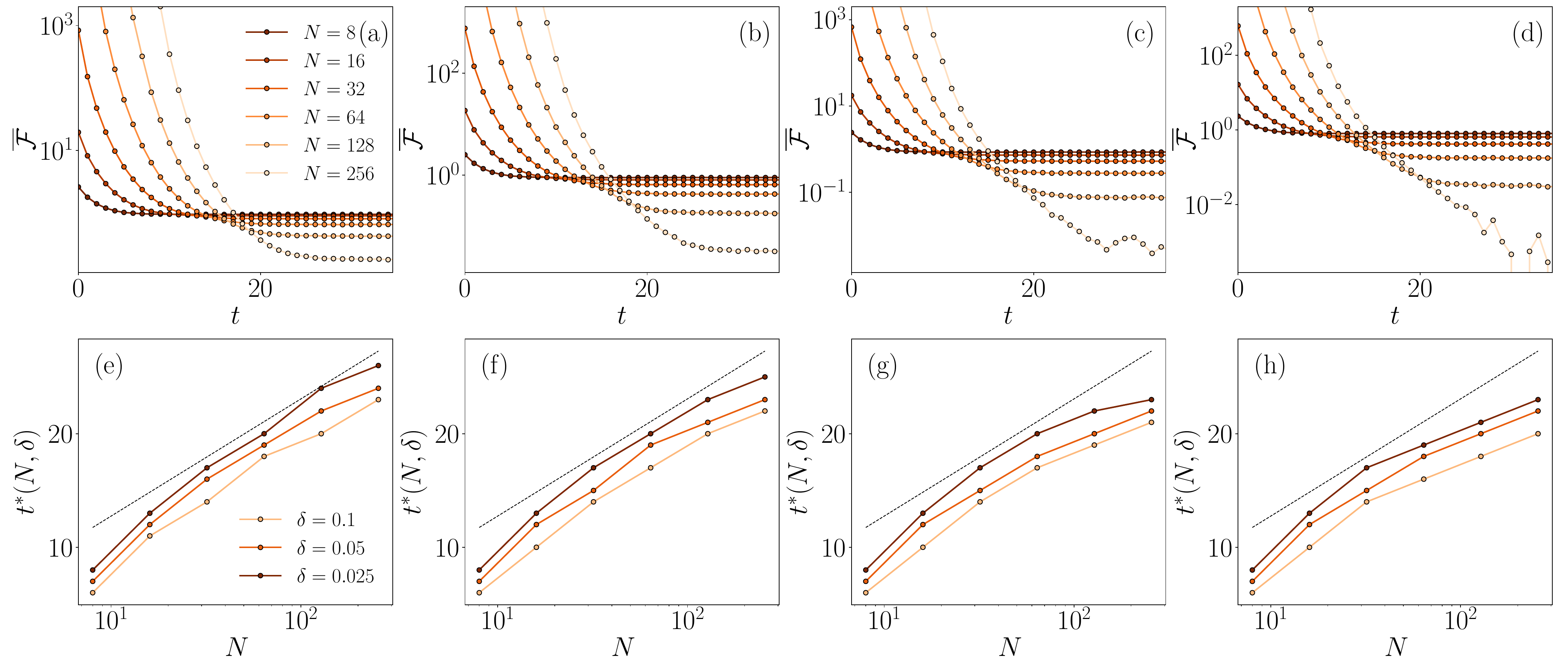}
\caption{Numerical simulation of the certification protocol for 1D local circuits. The target state $\ket{\psi}$ is the GHZ state, the lab state $\rho=\mathcal{E}_p^{\otimes N}(\ket{\psi}\bra{\psi})$, where $\mathcal{E}_p$ is the depolarizing channel. Each column represents a different value of $p$, increasing linearly from $p=0.01$ to $p=0.04$. (a)-(d): Averaged fidelity estimator $\overline{\mathcal{F}}=\mathbb{E}[\tilde{\mathcal{F}}]$ as a function of the circuit depth, for increasing $N$. (e)-(h): Depth $t^{\ast}(N,\delta)$ after which $\Delta F_{\psi,\rho} \leq \delta$.}
\label{fig:C2}
\end{figure*}

\begin{figure*}[t]
\includegraphics[trim={0.45cm 0.6cm 0.5cm 0},clip,width=\linewidth]{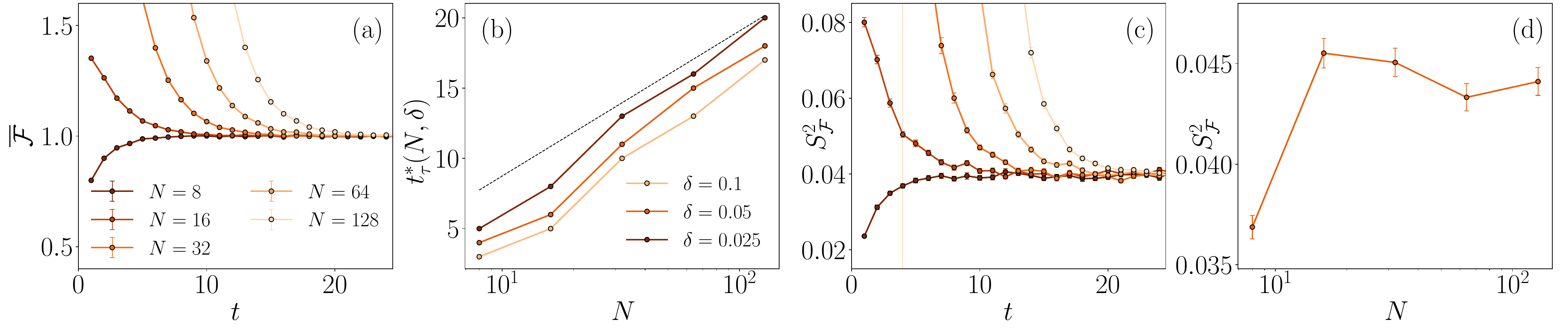}
\caption{Numerical simulation of the certification protocol for 1D local circuits, where both the lab state and the target state are equal to a single instance of a random short-range entangled stabilizer state. (a): Averaged fidelity estimator $\overline{\mathcal{F}}=\mathbb{E}[\tilde{\mathcal{F}}]$ as a function of the circuit depth, for increasing $N$. (b): Depth $t^{\ast}_{\tau}(N,\delta)$ after which $\Delta F_{\psi, \psi}\leq \delta$, where $\tau=3$ for the chosen state. The dashed line is the function in Eq.~\eqref{eq:t_star_bound}, shifted by a negative constant. (c): Variance of fidelity estimator $S^{2}_\mathcal{F}$ as a function of time. (d): The variance $S^{2}_\mathcal{F}$ estimated at the depth $t^\ast(N,\delta)$, for $\delta=0.05$.}
\label{fig:C3}
\end{figure*}

In this appendix, we provide details on the numerical methods employed in our work and additional data.

Stabilizer circuit simulations were employed for 1D states, both pure (GHZ state and random short-range entangled states) and mixed (through the application of depolarizing channels), and 2D states (cluster state). For this purpose, we used the \textsc{stim} library~\cite{gidney2021stim}.

All our results are obtained by constructing a suitable initial state and extracting $M = 50$ classical shadows with the shallow-circuit protocols described in Sec.~\ref{sec:standard_quantum_shadows}, for increasing circuit depth. The shadows are then post-processed to construct a single realization of our modified estimator $\tilde{\mathcal{F}}^{(s)}(t)$ for the fidelity. We then average over $R = 9600$ independent realizations of this estimator to obtain $\bar{\mathcal{F}}(t)$, and compute the sample variance $S_{\mathcal{F}}^2(t) = \sum_{s}(\tilde{\mathcal{F}}^{(s)}(t) - \bar{\mathcal{F}}(t))^2/R$. The error on the mean is then $\delta \mathcal{F}(t) = \sqrt{S_{\mathcal{F}}^2(t)/R}$. As for the sample variance, we construct 100 bootstrap samples and compute the variance over each of them. The standard deviation of the set of bootstrap variances is then used as the error on $S_{\mathcal{F}}^2(t)$.

In Fig.~\ref{fig:C1} we show two plots regarding non-local quantum circuits (extending Fig.~\ref{fig:fig3}(c) and Fig.~\ref{fig:fig3}(d) of the main text), once again in the case where both the lab state and the target state are the GHZ state. Fig.~\ref{fig:C1}(a) shows the values $t^{\ast}(N,\delta)$ after which $\Delta F_{\psi, \psi}\leq \delta$ for different values of $\delta$. In agreement with the bounds found in Ref.~\cite{dalzell2022random}, for this circuit architecture the scaling is also $a^{-1}\log(N)$, albeit with a prefactor which is smaller than the one in the local circuits counterpart. In Fig.~\ref{fig:C1}(b) we plot the variance $S_{\mathcal{F}}^2(t)$ at $t^{\ast}(N,\delta)$ for $\delta=0.05$, showing that it is bounded by a number independent of the system's size.

Going back to 1D local circuits, we consider again the case where the lab state is $\rho = \mathcal{E}_p^{\otimes N}(\ket{\psi}\bra{\psi})$, with $\ket{\psi}$ being the GHZ state. In Figs.~\ref{fig:C2}(a)-(d) we extend the plot of Fig.~\ref{fig:fig2}(a), showing the averaged fidelity estimator $\bar{\mathcal{F}} = \mathbb{E}[\tilde{\mathcal{F}}]$ as a function of the circuit depth, for increasing N, and for increasing depolarizing probability $p$. All plots share the same behaviour, meaning that the convergence of the fidelity estimator to its true value does not depend on the amount of noise affecting the lab state. This is also shown in Figs.~\ref{fig:C2}(e)-(h), where we plot the depth $t^{\ast}(N,\delta)$ after which $\Delta F_{\psi,\rho} \leq \delta$. We note that for increasing noise (and thus lower purities) the convergence is obtained at earlier depths.

Finally, we consider the case where the lab state $\ket{\psi}$ (and the target state) is a random short-range entangled stabilizer state, obtained by evolving the $\ket{0}$ state via a depth-3 shallow circuit $\ket{\psi} = U_3\ket{0}$ where all the two-qubit Clifford unitaries are equal to the same random two-qubit Clifford. In Fig.~\ref{fig:C3} we show the results when the latter is $(Z \otimes Z)\mathrm{CNOT}_{1,0}(\openone \otimes H)\mathrm{CNOT}_{0,1}(HS \otimes S)\mathrm{SWAP}$. We plot once again the averaged fidelity (a) and variance (c) as functions of circuit depth for increasing $N$. Fig.~\ref{fig:C3}(b) displays the depth $t^{\ast}_{\tau}(N,\delta)$ after which $\Delta F_{\psi, \psi} \leq \delta$, where $t^{\ast}_{\tau}(N,\delta) = t^{\ast}_0(N,\delta) - \tau$ as implied by Eq.~\ref{eq:w_circuit_final}. Comparing this with Fig.~\ref{fig:fig2}, we observe the same behaviour up to a shift of a constant $\tau = 3$, finding agreement with the analytical arguments presented in the main text. Finally, Fig.~\ref{fig:C3}(d) shows the variance $S_{\mathcal{F}^2}$ at depth $t^{\ast}_{\tau}(N,\delta)$, for $\delta = 0.05$, which is again bounded by a number independent of $N$.


\section{Details on the approximate purity estimators}
\label{sec:details_purity}

In this section we provide further detail on the computation of the approximate purity estimator.

\subsection{Mapping of the average estimators onto non-unitary dynamics}

We first explain the method to compute the average of the approximate estimator, which is based on a standard replica formalism~\cite{fisher2022random,potter2022entanglement}.

Consider the expectation value of the purity estimator for an initial state $\rho$:
\begin{equation}\label{eq:purity_average_formal}
    \mathbb{E}\left[\tilde{\mathcal{P}}_t^{(e)}\right] = 2^{2N} (2^N+1)^2 \int d\mu(U_t) d\mu(V_t) |\braket{0|U_t V_t^{\dagger}|0}|^2 \braket{0|U_t\rho U_t^{\dagger}|0} \braket{0|V_t\rho V_t^{\dagger}|0} - 2^N -2\,.
\end{equation}
It is convenient to view operators as states in a doubled Hilbert space, $\rho \rightarrow |\rho\rrangle$, so that we can identify $U\rho U^{\dagger} \rightarrow U\otimes U^{\ast} |\rho\rrangle$. The term inside the integral can be rewritten as the expectation value of the four-replicas operator $O = |\rho\rrangle \llangle\rho| \otimes I_{2^{2N}}$, on the state given by 
\begin{equation}
\mathbb{E}_{U_t}\left[U_t \otimes U_t^{\ast} \otimes U_t \otimes U_t^{\ast}  \right] |0^{\otimes 4}\rrangle\,.    
\end{equation}
Since $U_t$ is a quantum circuit where each gate is drawn randomly and independently, we can take the average over each gate individually see Ref.~\cite{fisher2022random} for more detail. In turn, since random Cliffords form a two-design, these averages can be performed analytically. Denoting by $U_{j,k}$ the two-qubit gate, we get
 \begin{align}\label{eq:two_qubit_gate}
 \notag \mathcal{U}_{jk} &=\mathbb{E}_{U_{j,k}}\left[U_{j,k}^{} \otimes U^{\ast}_{j,k} \otimes U_{j,k}^{} \otimes U_{j,k}^{\ast} \right] \\
 &= \frac{1}{d^4-1} \left[ |\mathcal{I}_{jk}^+\rrangle\llangle \mathcal{I}_{jk}^+| + |\mathcal{I}_{jk}^-\rrangle\llangle\mathcal{I}_{jk}^-| - \frac{1}{d^2}\left(|\mathcal{I}_{jk}^+\rrangle\llangle \mathcal{I}_{jk}^-| + |\mathcal{I}_{jk}^- \rrangle\llangle \mathcal{I}_{jk}^+| \right)\right]\,,
 \end{align}
  where $|\mathcal{I}_{jk}^{\pm}\rrangle = |\mathcal{I}^{\pm}\rrangle_j \otimes |\mathcal{I}^{\pm}\rrangle_k$, and the superstates $|\mathcal{I}^{\pm}\rrangle$ act as $\llangle A|\mathcal{I}^+\rrangle = \mathrm{Tr}(A S_{1,2} S_{3,4})$ and $\llangle A|\mathcal{I}^-\rrangle = \mathrm{Tr}(A S_{1,4} S_{2,3})$ for any four-replicas operator $A$, where $S_{a,b}$ is the two-replicas swap operator.

In summary, computing the average~\eqref{eq:purity_average_formal} requires evolving an initial state $|{0}\rrangle^{\otimes 4}$ by a non-unitary circuit made of the gates $\mathcal{U}_{j,k}$. Importantly, we see that $\mathcal{U}_{j,k}$ projects onto the local sub-subspace spanned by $|\mathcal{I}_{jk}^{\pm}\rrangle$. Therefore, the dynamics takes place in an effective chain of $N$ two-level systems.

For the 1D architecture, we have computed the evolved state using standard TN computations, using the fact that each layer of gates can be represented trivially as a matrix-product operator. Our implementation is in the \textsc{ITensor} library~\cite{itensor,itensor-r0.3}. While we find that the bipartite entanglement entropy grows very slowly with the number of layers, the contraction of tensors produces very small numerical values, so that the intermediate computations must be carried out at high numerical precision (recall that the dynamics is not unitary and does not preserve norms). This effectively limits the system sizes we can study to $N\lesssim 32$. In the case of non-local circuits, we can use a numerically exact method for larger sizes, as we detail in the following. 

\subsection{Averaged estimator for non-local circuits}
\label{sec:appendix_nonlocal}

For the non-local architecture considered in this work, $U_t$ is a non-local quantum circuit composed of $t$ ``layers" $U^{(j)}$, where each layer is composed of 2-qubit Clifford unitaries acting on $N/2$ pairs of qubits chosen at random, $U^{(j)} = U_{j_1 j_2} ... U_{j_{N-1} j_N}$. Averaging over the random pairs and the unitaries, we have

 \begin{equation}
    \mathbb{E}_{U^{(j)}}\left[U^{(j)} \otimes (U^{(j)})^{\ast} \otimes U^{(j)} \otimes (U^{(j)})^{\ast} \right]^t|0^{\otimes 4}\rrangle = \left( \frac{1}{N!}\sum_{j \in \mathrm{S}_N} \mathcal{U}_{j_1 j_2} ... \mathcal{U}_{j_{N-1}j_N} \right)^t |0^{\otimes 4}\rrangle = \mathcal{L}^t |0^{\otimes 4}\rrangle,
 \end{equation}
where $\mathcal{U}_{jk}$ is given in Eq.~\eqref{eq:two_qubit_gate}, while $\mathrm{S}_{N}$ is the permutation group. Thus the expectation value reads
 \begin{equation}
     \mathbb{E}\left[\tilde{\mathcal{P}}_t^{(e)}\right] = 2^{2N} (2^N+1)^2 \llangle 0^{\otimes 4}| \mathcal{L}^t O \mathcal{L}^t |0^{\otimes 4}\rrangle - 2^N - 2\,.
 \end{equation}
 To compute it, we first note that the $\mathcal{L}$ super-operator is a projector onto the Hilbert space spanned by $\{|\mathcal{I}^+\rrangle, |\mathcal{I}^-\rrangle \}^{\otimes N}$, that is, onto an $N$-fold tensor product of 2-dimensional Hilbert spaces. Since $\llangle \mathcal{I}^+|\mathcal{I}^+\rrangle = \llangle \mathcal{I}^-|\mathcal{I}^-\rrangle = d^2$ and $\llangle \mathcal{I}^+|\mathcal{I}^-\rrangle = d$, we can construct an orthonormal basis $\{|\tilde{0}\rrangle, |\tilde{1}\rrangle\}$ for each local Hilbert space, such that
 \begin{equation}
     |\mathcal{I}^+\rrangle = d|\tilde{0}\rrangle, \qquad |\mathcal{I}^-\rrangle = |\tilde{0}\rrangle + \sqrt{d^2-1}|\tilde{1}\rrangle\,.
 \end{equation}
 Next, we note that both the initial state $|0^{\otimes 4}\rangle$ and the super-operators $\mathcal{L}$ and $O$ are invariant under permutation of any of the $N$ sites. Thus we can construct a basis $\{|W_n\rrangle\}_{n=0}^{N}$ for the $N+1$-dimensional permutation invariant subspace:
 \begin{equation}
     |W_n\rrangle = \frac{1}{\sqrt{N!(N-n)!n!}}\sum_{\pi } \pi | \tilde{1}\rrangle^{\otimes n} |\tilde{0}\rrangle^{\otimes N-n}\,.
 \end{equation}
 Here and in the following, we denote by $\pi$ an arbitrary permutation operator over the $N$ two-level systems. Note that for convenience this basis can also be written in terms of the $\{|\mathcal{I}^+\rrangle, |\mathcal{I}^-\rrangle \}$ basis, as 
 \begin{equation}
     |W_n\rrangle = \frac{1}{\sqrt{N!(N-n)!n!}} \frac{1}{d^N (d^2-1)^{\frac{n}{2}}} \sum_{\pi} \left(d|\mathcal{I}^-\rrangle - |\mathcal{I}^+\rrangle\right)^{\otimes n} |\mathcal{I}^+\rrangle^{\otimes N-n} = \mathcal{N}^{-1}\sum_{\pi} |v\rrangle^{\otimes n} |\mathcal{I}^+\rrangle^{\otimes N-n},
 \end{equation}
 where we introduced $|v\rrangle = d|\mathcal{I}^-\rrangle - |\mathcal{I}^+\rrangle$.

 We are now ready to compute $\mathcal{L}|W_n\rrangle$. Fix the state $|v\rrangle^{\otimes n} |\mathcal{I}^+\rrangle^{\otimes N-n}$.  The action of each 2-qubit channel $\mathcal{U}_{jk}$ depends on whether the qubits $j,k$ are both in the $|v\rrangle$ state, both in the $|\mathcal{I}^+\rrangle$ state, or in opposite states. For a given sequence $\mathcal{U}_{j_1j_2}\ldots \mathcal{U}_{j_{N-1}j_N}$ (which identifies $N/2$ pairs), we denote the number of pairs in each of the three configurations by $r\leq n/2$, $s\leq (N-n)/2$ and $N/2-s-r$ respectively. Averaging over all permutations to obtain $\mathcal{L}$, we obtain
 \begin{align}
    \notag
    \mathcal{L}\sum_{\pi} \pi|v\rrangle^{\otimes n}|\mathcal{I}^+\rrangle^{\otimes N-n} ={}&
     \frac{1}{N!}\sum_{\pi} \pi\sum_{j \in \mathrm{S}_N} \frac{(d^2-1)^{N/2-s-r}}{(d^4-1)^{N/2-s}}\left[|v\rrangle^{\otimes 2} + 2|v\rrangle|\mathcal{I}^+\rrangle\right]^{\otimes r} \\
     \notag
     & \times |\mathcal{I}^+\rrangle^{\otimes 2s} \left[|v\rrangle^{\otimes 2} + 2|v\rrangle|\mathcal{I}^+\rrangle\right]^{\otimes N/2-s-r} \\
     \notag
     ={}& \sum_{\pi} \pi \frac{1}{N!}\sum_{j \in \mathrm{S}_N} \frac{(d^2-1)^{n}}{(d^4-1)^{n-r}}\left[|v\rrangle^{\otimes 2} + 2|v\rrangle|\mathcal{I}^+\rrangle\right]^{\otimes r} \\
     & \times |\mathcal{I}^+\rrangle^{\otimes N-2n+2r} \left[|v\rrangle^{\otimes 2} + 2|v\rrangle|\mathcal{I}^+\rrangle\right]^{\otimes n-2r}\,,
\end{align}
 where $r$ and $s$ depend on permutation $j$. Note that in the second line we used the fact that $r$ and $s$ are related by $n-2r = N-n-2s$, and thus $s=N/2-n+r$. Now we can expand the tensor products and reorder each term:
 \begin{equation}
 \sum_{\pi}  \pi \sum_{j \in \mathrm{S}_N} \left[|v\rrangle^{\otimes 2} + 2|v\rrangle|\mathcal{I}^+\rrangle\right]^{\otimes r} =  \sum_{\pi} \pi \sum_{j \in \mathrm{S}_N} \sum_{k=0}^r \binom{r}{k} |v\rrangle^{\otimes r+k} |\mathcal{I}^+\rrangle^{\otimes r-k} 2^{r-k},
 \end{equation}
 and analogously for $\left[|v\rrangle^{\otimes 2} + 2|v\rrangle|\mathcal{I}^+\rrangle\right]^{\otimes n-2r}$. We arrive at
\begin{align}
\notag
   \mathcal{L}  \sum_{\pi} \pi|v\rrangle^{\otimes n}|\mathcal{I}^+\rrangle^{\otimes N-n}
    ={}& \frac{(d^2-1)^n}{N!}\sum_{\pi} \pi\sum_{j \in \mathrm{S}_N} \sum_{k=0}^r \sum_{l=0}^{n-2r} \frac{2^{n-r-(k+l)}}{(d^4-1)^{n-r}} \binom{r}{k}\binom{n-2r}{l} \\ 
    & \times |v\rrangle^{\otimes n-r+(k+l)} |\mathcal{I}^+\rrangle^{\otimes N-n+r-(k+l)}\,.
\end{align}
Next, we set $m=n-r+(k+l)$, yielding
\begin{align}
\sum_{k=0}^r \sum_{l=0}^{n-2r} \frac{2^{n-r-(k+l)}}{(d^4-1)^{n-r}} \binom{r}{k}\binom{n-2r}{l} |v\rrangle^{\otimes n-r+(k+l)} |\mathcal{I}^+\rrangle^{\otimes N-n+r-(k+l)}=
     \sum_{m=0}^N c_m^{(r)} |v\rrangle^{\otimes m}|\mathcal{I}^+\rrangle^{\otimes N-m}\,,
\end{align}
where, using the Vandermonde's identity,
\begin{equation}
    c_m^{(r)} = \sum_{k=0}^r \sum_{l=0}^{n-2r} \frac{2^{n-r-(k+l)}}{(d^4-1)^{n-r}} \binom{r}{k}\binom{n-2r}{l} \delta_{m,n-r+(k+l)} = \frac{2^{2(n-r)-m}}{(d^4-1)^{n-r}}\binom{n-r}{m-n+r}\,.
\end{equation}
Therefore
\begin{align}
    \notag
   \mathcal{L}  \sum_{\pi} \pi|v\rrangle^{\otimes n}|\mathcal{I}^+\rrangle^{\otimes N-n}
    &= \frac{(d^2-1)^n}{N!}\sum_{\pi}\pi  \sum_{j \in \mathrm{S}_N} \sum_{m=0}^N c_m^{(r)} |v\rrangle^{\otimes m}|\mathcal{I}^+\rrangle^{\otimes N-m} \\
    \notag
    &= \frac{(d^2-1)^n}{N!} \sum_{j \in \mathrm{S}_N} \sum_{m=0}^{N} c_m^{(r)} \sum_{\pi} |v\rrangle^{\otimes m}|\mathcal{I}^+\rrangle^{\otimes N-m} \\ 
    &= \frac{(d^2-1)^n}{N!} \sum_{j \in \mathrm{S}_N} \sum_{m=0}^N (N!(N-m)! m!)^{\frac{1}{2}} d^N(d^2-1)^{\frac{m}{2}}c_{m}^{(r)}|W_m\rrangle\,.
\end{align}

We can now perform the sum over permutations $j \in \mathrm{S}_N$ by counting the number of permutations $N_r$ with fixed $r$, and then summing over $r$. The reasoning is the following. We have $N/2$ pairs divided in the following way: $r$ pairs of type $|v\rrangle^{\otimes 2}$, $s$ pairs of type $|\mathcal{I}^+\rrangle^{\otimes N-n}$, and $N/2-s-r$ pairs of type $|v\rrangle|\mathcal{I}\rrangle$ or $|\mathcal{I}\rrangle|v\rrangle$. Thus we have $\binom{N/2}{r}\binom{N/2-r}{s}$ different ways to order them. We are then free to permute the indices $j_1...j_N$ in a way that preserves this order. Thus there are $n!$ permutations of the $|v\rrangle$ states, $(N-n)!$ permutations of the $|\mathcal{I}^+\rrangle$ states, and $2^{N/2-s-r}$ ways to order the states in the mixed pairs. Knowing that $s = N/2-n+r$, the total is 
\begin{equation}
    N_r = \binom{N/2}{r}\binom{N/2-r}{N/2-n+r}n!(N-n)!2^{n-2r} = \binom{N/2}{r, n-2r, N/2-n+r}n!(N-n)!2^{n-2r}\,.
\end{equation}
Summing over $r$, we arrive at
\begin{align}
\notag
    \mathcal{L}\sum_{\pi}\pi |v\rrangle^{\otimes n}|\mathcal{I}^+\rrangle^{\otimes N-n} = \frac{d^N(d^2-1)^n}{\binom{N}{n}} \sum_{m=0}^N (d^2-1)^{\frac{m}{2}}\sqrt{N!m!(N-m)!} |W_m\rrangle \\
    \times \sum_{r=0}^{n/2} \binom{N/2}{r, n-2r, N/2-n+r}2^{n-2r} c_{m}^{(r)}
\end{align}
and thus, finally
\begin{equation}
    \mathcal{L}|W_n\rrangle = \frac{(d^2-1)^{\frac{n}{2}}}{\binom{N}{n}} \sum_{m=0}^N (d^2-1)^{\frac{m}{2}} \sqrt{\frac{m!(N-m)!}{n!(N-n)!}} d^{(n)}_m |W_m\rrangle\,,
\end{equation}
where 
\begin{equation}
 d^{(n)}_m=\sum_{r=0}^{n/2} \binom{N/2}{r, n-2r, N/2-n+r}2^{n-2r} c_{m}^{(r)}\,.
\end{equation}

This expression can now be used to numerically compute $\mathcal{L}^t$ in the permutation-invariant basis $\{|W_n\rrangle\}$. This makes the computation efficient, because the corresponding Hilbert space only grows linearly in $N$. It only remains to expand both the initial state $|0^{\otimes 4}\rrangle$ and the super-operator $O = |\rho\rrangle\llangle\rho| \otimes I_{2^{2N}}$ in the permutation invariant basis $\{|W_n\rrangle\}_{n=0}^N$.

The initial state is a product state $|0^{\otimes 4}\rrangle = \bigotimes_{j=1}^N |0^{\otimes 4}\rrangle_j$, so we can compute its projection on the local basis $\{|\tilde{0}\rrangle, |\tilde{1}\rrangle\}$ to see that
\begin{equation}
    |0^{\otimes 4}\rrangle = \frac{1}{d^N} \left(|\tilde{0}\rrangle + \sqrt{\frac{d-1}{d+1}}|\tilde{1}\rrangle\right)^{\otimes N} = \frac{1}{d^N} \sum_{n=0}^{N} \left(\frac{d-1}{d+1}\right)^{\frac{n}{2}} \binom{N}{n}^{\frac{1}{2}} |W_n\rrangle\,.
\end{equation}

For the super-operator, we focus on the case where $\rho$ is a product state, $\rho = \bigotimes_{j=1}^N \rho_j$. Then we can compute the matrix elements of $O_j = \rho_j \otimes I_{4}$ over $\{|\tilde{0}\rrangle, |\tilde{1}\rrangle\}_j$, and obtain
\begin{equation}
    O_j |\tilde{0}\rrangle_j = \frac{1}{d}|\tilde{0}\rrangle_j, \qquad O_j |\tilde{1}\rrangle_j = \frac{1}{d^2-1}\left(\mathrm{tr}\rho_j^2 - \frac{1}{d} \right)|\tilde{1}\rrangle_j,
\end{equation}
and thus
\begin{equation}
    \llangle W_m | O | W_n\rrangle = \delta_{m,n} \frac{1}{d^N} \frac{1}{(d^2-1)^n} \binom{N}{n}^{-1} \sum_{j \in \tilde{\pi}} \left(d\mathrm{tr}\rho_{j_1}^2 - 1 \right)...\left(d\mathrm{tr}\rho_{j_n}^2 - 1 \right) 1_{j_{n+1}...j_N},
\end{equation}
where the sum is over all the different subsets of $n$ elements $\{\rho_{j_1}, ..., \rho_{j_n}\}_j$ of the set of single-qubit states $\{\rho_k\}_{k=1}^N$. If the state is translational invariant, the expression finally simplifies to
\begin{equation}
    \llangle W_m | O | W_n\rrangle = \delta_{m,n} \frac{1}{d^N} \frac{1}{(d^2-1)^n} \left(d\mathrm{tr}\rho_{j}^2-1\right)^n\,.
\end{equation}

\subsection{Variance of the global Clifford purity estimator}
\label{sec:appendix_purity_variance}

We want to compute the variance of the purity estimator $\tilde{\mathcal{P}}_t^{(e)}$ defined in Eq.~\eqref{eq:approximated_estimator_purity} when $t \to \infty$, that is, when our estimator reduces to the global Clifford estimator. Following \cite{huang2020predicting}, we have
\begin{align}\label{eq:D3var}
    \notag
    \mathrm{Var}\left[\tilde{\mathcal{P}}_{\infty}^{(e)} \right] &= \binom{M}{2}^{-2} \sum_{r< r'}\sum_{s<s'} \left( \mathbb{E}\left[ \mathrm{Tr}(\tilde{\rho}_{\infty}^{(r)}\tilde{\rho}_{\infty}^{(r')})\mathrm{Tr}(\tilde{\rho}_{\infty}^{(s)}\tilde{\rho}_{\infty}^{(s')}) \right] - \mathcal{P}(\rho)^2 \right) \\
    &= \binom{M}{2}^{-1}\mathrm{Var}\left[\mathrm{Tr}(\tilde{\rho}_{\infty}^{(1)} \tilde{\rho}_{\infty}^{(2)})\right] + \binom{M}{2}^{-1}2(M-2) \mathrm{Var}\left[\mathrm{Tr}(\tilde{\rho}_{\infty}^{(1)} \rho) \right]\,.
\end{align}
The first term can be written explicitly as
\begin{align}
    \notag
    \mathrm{Var}\left[\mathrm{Tr}(\tilde{\rho}_{\infty}^{(1)} \tilde{\rho}_{\infty}^{(2)})\right] ={}& \underset{U,V \in \mathrm{Cl}(2^N)}{\mathbb{E}} \underset{b,d}{\mathbb{E}} \ \mathrm{Tr}\left[\left((2^N+1)U^{\dagger}|b\rangle\langle b|U - \openone \right) \left((2^N+1)V^{\dagger}|d\rangle\langle d|V - \openone \right) \right]^2 - \mathcal{P}(\rho)^2 \\
    \notag
    ={}& \int d\mu_H(U) d\mu_H(V) \left\{ 2^{2N}\left[(2^N+1)^2 |\braket{0|UV^{\dagger}|0}|^2 - (2^N+2) \right]^2 \right. \\
    \notag
    & \left. \times \braket{0|U\rho U^{\dagger}|0}\braket{0|V\rho V^{\dagger}|0} \right\} - \mathcal{P}(\rho)^2 \\
    \notag
    ={}& 2^{2N}(2^N+1)^4 \int d\mu_H(U) d\mu_H(V) |\braket{0|UV^{\dagger}|0}|^4 \braket{0|U\rho U^{\dagger}|0}\braket{0|V\rho V^{\dagger}|0}\\ 
    & - (2^N + 2 + \mathcal{P}(\rho))^2 \,,
\end{align}
where in the first line we used the fact that for $t \to \infty$ the random circuit in our estimator reduces to a random global Clifford unitary $U \in \mathrm{Cl}(2^N)$, and in the second line we invoked the 3-design property of Clifford unitaries to replace the sum with an integral over Haar measure $d\mu_H(U)$.
The first term in the last line can then be rewritten as
\begin{equation}
    2^{2N}(2^N+1)^4 \int d\mu_H(U) \braket{0|U\rho U^{\dagger}|0} \mathrm{Tr}\left\{ \left((U^{\dagger})^{\otimes 2}|0\rangle\langle 0|^{\otimes 2}U^{\otimes 2} \otimes \rho \right) \int d\mu_H(V) (V^{\dagger})^{\otimes 3}|0\rangle \langle 0|^{\otimes 3} V^{\otimes 3}\right\},
\end{equation}
and now the last integral can be computed using~\eqref{eq:haar3}, yielding
\begin{align}
    \notag
    &= \frac{2^N (2^N+1)^3}{2^N+2} \int d\mu_H(U) \braket{0|U\rho U^{\dagger}|0} (2 + 4\braket{0|U\rho U^{\dagger}|0}) \\
    &= \frac{2^N (2^N+1)^3}{2^N+2} \left(\frac{2}{2^N} + 4\frac{1+\mathcal{P}(\rho)}{2^N(2^N+1)} \right) = \frac{2(2^N+1)^2}{2^N+2}\left(2^N+3 + 2\mathcal{P}(\rho) \right),
\end{align}
thus arriving at
\begin{equation}
    \mathrm{Var}\left[\mathrm{Tr}(\tilde{\rho}_{\infty}^{(1)} \tilde{\rho}_{\infty}^{(2)})\right] = \frac{2(2^N+1)^2}{2^N+2}\left(2^N+3 + 2\mathcal{P}(\rho) \right) - \left(2^N + 2 + \mathcal{P}(\rho) \right)^2\,.
\end{equation}
Following similar steps, the second term in Eq.~\eqref{eq:D3var} can be written as
\begin{align}
    \notag
    \mathrm{Var}\left[\mathrm{Tr}(\tilde{\rho}_{\infty}^{(1)} \rho) \right] &= \underset{U \in \mathrm{Cl}(2^N)}{\mathbb{E}} \underset{b}{\mathbb{E}} \ \mathrm{Tr}\left[\left((2^N+1)U^{\dagger}|b\rangle\langle b|U - \openone \right)\rho \right]^2 - \mathcal{P}(\rho)^2 \\
    \notag
    &= 2^N \int d\mu_H(U) \left((2^N+1)^2 \braket{0|U\rho U^{\dagger}|0}^2 -2(2^N+1)\braket{0|U\rho U^{\dagger}|0}+1\right)\braket{0|U\rho U^{\dagger}|0} - \mathcal{P}(\rho)^2 \\
    \notag
    &= \frac{2^N+1}{2^N+2}\left(1+3\mathcal{P}(\rho) + 2\mathrm{Tr}(\rho^3)\right) - 2\left(1+\mathcal{P}(\rho)\right) + 1 - \mathcal{P}(\rho)^2 \\
    &= \frac{2^N+1}{2^N+2}\left(1+3\mathcal{P}(\rho) + 2\mathrm{Tr}(\rho^3)\right) - \left( 1 + \mathcal{P}(\rho) \right)^2\,.
\end{align}
Putting everything together, we finally arrive at
\begin{align}\label{eq:variance_purity_exact}
    \notag
    \mathrm{Var}\left[\tilde{\mathcal{P}}_{\infty}^{(e)} \right] ={}& \binom{M}{2}^{-1}\left\{\frac{2(2^N+1)^2}{2^N+2}\left(2^N+3 + 2\mathcal{P}(\rho) \right) - \left(2^N + 2 + \mathcal{P}(\rho) \right)^2 \right. \\
    & + \left. 2(M-2)\left[\frac{2^N+1}{2^N+2}\left(1+3\mathcal{P}(\rho) + 2\mathrm{Tr}(\rho^3)\right) - \left(1+\mathcal{P}(\rho) \right)^2\right] \right\}\,.
\end{align}
Thus the leading term is $\mathrm{Var}[\tilde{\mathcal{P}}_{\infty}^{(e)}] \sim \frac{2(2^N)}{M(M-1)}$.
\newpage
\end{widetext}

\bibliographystyle{quantum}
\bibliography{bibliography}

\end{document}